\documentclass[12pt,a4paper,twoside]{article}
\usepackage[top=1in, bottom=1.in, left=1in, right=1in]{geometry}
\usepackage[title,toc,page]{appendix} 
\usepackage{titlesec}
\usepackage[utf8]{inputenc} 
\usepackage[T1]{fontenc} 
\usepackage{fancyhdr}
\usepackage{soul}

\fancypagestyle{Fancypages}{%
    \fancyhf{}%

    \fancyhead[LE,RO]{{\sc Page \thepage}}
    \fancyhead[RE,LO]{ }
    \fancyfoot[CE,CO]{ }
    \fancyfoot[LE,RO]{ }}

\fancypagestyle{Fancytitle}{%
    \fancyhf{}%

    \fancyfoot[LE,RO]{ }}

\usepackage{authblk}
\usepackage[colorlinks,citecolor=blue,linkcolor=blue]{hyperref}
\usepackage{lscape}

\usepackage{mathtools}
\mathtoolsset{showonlyrefs}
\usepackage{amsthm}
\usepackage{mathrsfs}
\usepackage{amsmath}
\usepackage{amsfonts}
\usepackage{amssymb}
\usepackage{yfonts}
\usepackage{BOONDOX-calo}
\usepackage{enumerate}
\usepackage{booktabs}
\usepackage{multirow}

\usepackage[dvipsnames]{xcolor}

\usepackage{tikz}

\titleformat{\section}
{\normalfont\fontsize{16}{15}\scshape}{\thesection}{1em}{}

\titleformat{\subsection}
{\normalfont\fontsize{14}{15}\scshape}{\thesubsection}{1em}{}

\titleformat{\subsubsection}
{\normalfont\fontsize{13}{15}\scshape}{\thesubsubsection}{1em}{}

\renewenvironment{abstract}{%
\hfill\begin{minipage}{0.95\textwidth}
\rule{\textwidth}{1pt}}
{\par\noindent\rule{\textwidth}{1pt}\end{minipage}}

\newtheorem{remark}{Remark}[section]
\newtheorem{Def}{\sc Definition}[section]
\newtheorem{example}{\sc Example}
\newtheorem{theorem}{\sc Theorem}[section]
\newtheorem{Prop}{\sc Proposition}[section]

\newtheorem{Ass}{Assumption}[section]

\usepackage{graphicx}
\usepackage{tikz-cd}
\usepackage{color}
\usetikzlibrary{arrows.meta}
\usetikzlibrary{positioning}

\usepackage[autostyle=true]{csquotes} 
\usepackage[backend=bibtex,natbib=false,style=numeric,sortcites=true]{biblatex} 
\renewbibmacro{in:}{}
\bibliography{quadratic.bib} 

\newcommand{\reeb}{\mathcal{R}} 
\newcommand{\cH}{\mathcal{H}} 

\newcommand{\ch}{\mathcal{h}}
\newcommand{\cO}{\mathcal{O}}
\DeclareMathOperator{\Span}{Span}
\newcommand{\PB}[2]{\left\{\,#1\,,\,#2\,\right\}_{PB}}

\newcommand{\orcid}[1]{\href{https://orcid.org/#1}{\textcolor[HTML]{A6CE39}{#1}}}
\usepackage{todonotes}

\title{\sc The flow method for the Baker-Campbell-Hausdorff formula: exact results}

\author[1]{Federico Zadra \footnote{\href{mailto:f.zadra@rug.nl}{f.zadra@rug.nl} \orcid{0000-0001-7565-3792}}}
\author[2]{Alessandro Bravetti \footnote{\href{mailto:alessandro.bravetti@iimas.unam.mx}{alessandro.bravetti@iimas.unam.mx} \orcid{0000-0001-5348-9215}}}
\author[3,4]{Angel Alejandro Garc\'ia-Chung \footnote{\href{mailto:alechung@tec.mx}{alechung@tec.mx} \orcid{0000-0002-8982-3569}}}
\author[1]{Marcello Seri \footnote{\href{mailto:m.seri@rug.nl}{m.seri@rug.nl} \orcid{0000-0002-8563-5894}}}

\affil[1]{Bernoulli Institute for Mathematics, Computer Science and Artificial Intelligence, University of Groningen, Groningen, The Netherlands}
\affil[2]{Instituto de Investigaciones en Matemáticas Aplicadas y en Sistemas (IIMAS--UNAM), Mexico City, Mexico}
\affil[3]{Tecnol\'ogico de Monterrey, Escuela de Ingenier\'ia y Ciencias, Carr. al Lago de Guadalupe Km. 3.5, Estado de Mexico 52926, Mexico.}
\affil[4]{Max Planck Institute for Mathematics in the Sciences Inselstraße 22, 04103 Leipzig, Germany}

\date{ }

\begin{document}

\maketitle

\thispagestyle{Fancytitle}
\begin{abstract}
  Leveraging techniques from the literature on geometric numerical integration,
  we propose a new general method to compute exact expressions for the BCH formula.
  In its utmost generality, the method consists in embedding the Lie algebra of interest
  into a subalgebra of the algebra of vector fields on some manifold by means of an isomorphism,
  so that the BCH formula for two elements of the original algebra can be recovered from the composition
  of the flows of the corresponding vector fields. For this reason we call our method
  the \emph{flow method}.
  Clearly, this method has great advantage in cases
  where the flows can be computed analytically. 
  We illustrate  
  its usefulness on some benchmark examples
  where it can be applied directly, and discuss
  some possible extensions for cases where an exact expression cannot be obtained.
\end{abstract}

\tableofcontents

\section*{Introduction}

The Baker-Campbell-Hausdorff (BCH) formula is a well known result in the theory of Lie groups and Lie algebras~\cite{Achilles2012, Hall2015, Fulton1999} that found wide application in both mathematics and physics: from Lie theory~\cite{Hall2015} to numerical integration~\cite{Bravetti2020,Skeel2001}, quantum mechanics~\cite{Bagarello_2022, Sakurai2011} and statistical mechanics~\cite{Tuckerman2010}. Roughly speaking, it links the composition between two elements of a Lie group in a connected neighbourhood of the identity to an element of the Lie algebra whose exponential map is the group element which results from the composition.
In other words, it provides a means to compute the map $Z:\mathfrak{g} \times\mathfrak{g} \to \mathfrak{g}$ defined by
\begin{equation}\label{eq:logXY}
    Z(A,B):= \log(e^A e^B)\,,
\end{equation}
where $e^A,e^B\in G$ are the corresponding elements in the associated Lie group given by the exponential map.
When the Lie algebra is commutative, the BCH formula is simply
\begin{equation}
  Z(A, B) = A + B.
\end{equation}
However, Lie algebras are in general not commutative and finding closed-form expressions or even accurate estimates for $Z$ can become a daunting task.
The general \emph{BCH formula} can be expressed (formally) in terms of 
the integral formulation
\begin{equation}\label{eq:integralBCH}
  Z(A,B) 
  = A + B - \int_0^1 \sum_{n=1}^\infty \frac{(\mathbb{I} - e^{ad_A} e^{t\,ad_B})^n}{n (n+1)}\,dt\,,
\end{equation}
where $ad_A$ is the adjoint representation of $A$ {and $\mathbb{I}$ is the identity matrix}.
To approximate~\eqref{eq:integralBCH}, the most common method is to employ the series expansion
\begin{equation}\label{eq:seriesBCH}
  Z(A,B) = A + B + [A,B] + \frac{1}{12} \left([A,[A,B]] + [B,[B,A]] \right) + \cdots
\end{equation}
For a detailed treatment of the topic, we refer the reader to~\cite{Varadarajan1974, Fulton1999, Hall2015}.

The problem of computing a closed form for~\eqref{eq:integralBCH} for a given particular algebra 
has already been addressed in various settings, employing a number of different algebraic and analytic 
techniques~\cite{Van_Brunt_2018, Matone_JGP_2015, Matone_JHEP_2015,Fulton1999,Weigert1997,Bravetti2017, Biagi2013a}.
In this paper, we present a dynamic approach to such problem: we compute closed forms of the BCH formula by exploiting 
an isomorphism between the given Lie algebra and an appropriate subalgebra of the infinitesimal contactomorphisms on a contact manifold. The latter in turn is also isomorphic to the Lie algebra of the Hamiltonian functions
endowed with the Jacobi bracket that naturally arises from the contact structure~\cite{de_Le_n_2019}.
This helps us to simplify the calculations.
The mapping between functions and vector fields is always an isomorphism for contact Hamiltonians~\cite{de_Le_n_2019}, differently from the symplectic case~\cite{Fasano2002}. Therefore, even in the case of some algebras which arise naturally in the symplectic setting (e.g.~the Heisenberg group), the use of the contact structure is more natural, as it guarantees a one-to-one correspondence between contact Hamiltonian functions and contact Hamiltonian vector fields.

We prove the usefulness of our method using some interesting examples arising from the recent literature in contact geometry and Hamiltonian systems, namely, the Heisenberg algebra~\cite{Hall2015}, the contact Heisenberg algebra~\cite{Bravetti2017}, the quadratic symplectic and the quadratic contact algebras, and {the complexification of $su(2)$}. Furthermore, we also include an analysis of contact splitting numerical integrators~\cite{Bravetti2020, Zadra_2021} for the damped harmonic oscillator~\cite{Bravetti2017a}. More precisely, in this example our method will be used in order to obtain
a priori estimates of the numerical errors for different splitting integrators.

The paper is organized as follows: in Section~\ref{sec:contactgeometry} we briefly review some concepts in 
contact geometry and contact Hamiltonian mechanics. In Section~\ref{sec:flow} we describe the flow method.
Section~\ref{sec:examples} is devoted to the aforementioned examples, presented in detail, 
while Section~\ref{sec:Conclusions} presents the conclusions and some additional directions for further research.
Additional results, including a matrix representation of the algebras treated in the paper, 
are presented at the end of the paper in Appendix~\ref{app:matrixrepr}.

\section{Contact geometry} \label{sec:contactgeometry}

We summarize here the essential concepts of contact geometry that are needed for this work and refer the reader to~\cite{P.Libermann1987,arnol2013mathematical,Boyer2011,Bravetti2017a, Bravetti2017b,de_Le_n_2019,gaset2020new,bravetti2022scaling,grabowska2022novel} 
for additional details, 
including the relevant proofs. 
The main defining object for us is a special differential $1$-form, called the \emph{contact form}, which induces both a volume form on the manifold and a maximally non-integrable distribution on its tangent bundle, called the \emph{contact structure}.
\begin{Def}\rm
  \label{def:contactmanifold}
  A (exact) \emph{contact manifold} is a couple $(M,\eta)$, where $M$ is $2n+1$-dimensional manifold and $\eta$ is a $1$-form such that
  \begin{equation}
    \eta \wedge (d\eta)^n \neq 0.
  \end{equation}
 In this case $\eta$ is called a \emph{contact form} and its kernel at each point defines a maximally non-integrable distribution of hyperplanes $\mathcal D$, called the \emph{contact structure}.
\end{Def}
If, on the same manifold $M$, the contact form $\eta$ is replaced by $\eta' = f \eta$, for some nowhere-vanishing real function $f$, the new $1$-form is again a contact form and defines the same contact structure.
Indeed, the induced volume form is simply
\begin{equation}
  \eta' \wedge (d \eta')^n = f^{n+1} \eta \wedge (d\eta)^n \neq 0.
\end{equation}
\pagestyle{Fancypages}
At each $p\in M$, the contact $1$-form $\eta_p$ and its differential $d\eta_p$ split the tangent space $T_pM$ into the so-called vertical and the horizontal distributions:
\begin{equation}
  \label{eqn:splittingTM}
  \mathcal D_p=\mathrm{Hor}_p(\eta) = \ker(\eta_p) , \qquad \mathrm{Vert}_p(\eta) = \ker(d\eta_p)\,,
\end{equation}
where $\ker(d\eta_p)$ is the subspace of $T_pM$ of all those vectors that annihilate $d\eta_p$.
Thus the tangent bundle $TM$ can be written as a Whitney sum $TM= \mathcal D_p\bigoplus \mathrm{Vert}_p(\eta)$.
It is important to stress that the definition of the horizontal distribution is invariant under the transformation $\eta \mapsto f \eta$, since $\ker(f\eta) = \ker(\eta)$, while the vertical distribution is not~\cite{P.Libermann1987}.
The vertical distribution is $1$-dimensional and is described in terms of the so-called Reeb vector field, which is unique once a contact form is fixed.
\begin{Prop}\label{prop:reeb}\rm
  Given a contact manifold $(M,\eta)$ there exists a unique vector field $\reeb$, the \emph{Reeb vector field}, such that
  \begin{equation}
    \label{eqn:rebbdef}
    \eta(\reeb) = 1 \qquad\text{and}\qquad d \eta(\reeb, \cdot) = 0\,.
  \end{equation}
\end{Prop}

Diffeomorphisms that preserve the contact structure are of particular interest. We have seen that two contact forms define the same contact structure if they are equal up to multiplication by a non-vanishing function: this allows to classify contactomorphisms into two classes.
\begin{Def}\rm
  \label{def:groupcontactomorphism}
  Denote $(M,\eta)$ and $(M', \eta')$ two contact manifolds which are diffeomorphic via $\phi : M \to M'$. We call the map $\phi$ a
  \begin{itemize}
    \item \emph{contactmorphism}, if there exists
    a non-vanishing
    $\lambda \in\mathcal{C}^\infty(M, \mathbb{R})$ such that $\phi^* \eta' = \lambda \eta$,
    \item \emph{strict contactomorphism}, if $\phi^* \eta' = \eta$.
  \end{itemize}
\end{Def}
Similarly, we can also classify the corresponding infinitesimal transformations.
\begin{Def}\rm
  \label{def:vectorcontactomorphism}
  Let $(M,\eta)$ be a contact manifold. A vector field $X$ on $M$ is called
  \begin{itemize}
    \item \emph{infinitesimal contactomorphism}, if $L_X \eta = \tau \eta$ for some $\tau \in \mathcal{C}^{\infty}(M,\mathbb{R})$,
    \item \emph{strict infinitesimal contactomorphism}, if $L_X \eta = 0$.
  \end{itemize}
\end{Def}

Finally, we want to give a coordinate description of a contact manifold.
In Definition~\ref{def:contactmanifold} we have seen that the distribution is described by a 1-form. The following theorem provides local canonical coordinates.

\begin{theorem}[Darboux's theorem]   \label{thm:darbouxtheorem}\rm
  Consider a contact manifold $(M,\eta)$ and a point $x \in M$.
  Then there exist local coordinates $(q_i,p_i,s)$ in a neighbourhood of $x$ such that the contact form is written as
  \begin{equation}
    \eta = ds - p_i dq_i,
  \end{equation}
where here and in the following Einstein's summation convention over repeated indices will be always assumed.
\end{theorem}
The coordinates in Darboux's theorem are called \emph{canonical coordinates}.  
Note that in these coordinates we have
$d \eta = d q_i \wedge d p_i$ and $\reeb = \frac{\partial}{\partial s}$.

\subsection{Contact hamiltonian flows and Jacobi structures}

The contact form allows to define an isomorphism between functions and vector fields. We call these vector fields contact gradients or contact Hamiltonian vector fields~\cite{P.Libermann1987}.
\begin{Def}\label{def:contactgradient}\rm
  Given a contact manifold $(M,\eta)$ and a smooth function $\cH : M \to \mathbb{R}$,
  the \emph{contact Hamiltonian vector field} associated with  $\cH$
  is the vector field $X_\cH \in \mathfrak{X}(M)$ such that
  \begin{equation}\label{eq:corrXH}
    \eta (X_\cH) = - \cH, \qquad
    d\eta(X_\cH, \cdot) = d\cH - d\cH(\reeb) \eta.
  \end{equation}
 We denote the map that assigns to a function $\cH$ its associated contact Hamiltonian vector field as $a_{\eta}$, 
 so that $X_{\cH}=a_{\eta}(\cH)$ and the function $\cH$ is called   \emph{contact Hamiltonian function}.
\end{Def}
Any contact Hamiltonian vector field is an infinitesimal contactomorphism, as shown by a direct application of Cartan's magic formula:
\begin{equation}
  \mathcal{L}_{X_\cH} \eta = d (\eta(X_\cH)) + d\eta (X_\cH, \cdot) = - \reeb(\cH) \eta.
\end{equation}
In particular, it is strict if and only if $\reeb(\cH) = 0$.
Remarkably, one can prove also the opposite, i.e.~any infinitesimal contactomorphism is a contact Hamiltonian vector field, with the corresponding (contact) Hamiltonian function being recovered by using the first condition in~\eqref{eq:corrXH}.

In canonical coordinates $(q_i,p_i,s)$, the contact Hamiltonian vector field is given by~\cite{Bravetti2017a,Bravetti2017b}
\begin{equation}
  X_\cH = \left(\frac{\partial \cH}{\partial p_j}\right) \frac{\partial}{\partial q_j} + \left(-\frac{\partial \cH}{\partial q_j} - p_j \frac{\partial \cH}{\partial s} \right) \frac{\partial}{\partial p_j} + \left(p_i \frac{\partial \cH}{\partial p_i} - \cH \right) \frac{\partial}{\partial s}
\end{equation}
and its integral curves satisfy the contact Hamiltonian equations of motion
\begin{equation}\label{eq:CHameqs}
  \begin{cases}
    \dot q_j = \frac{\partial \cH}{\partial p_j}                                        \\
    \dot p_j = -\frac{\partial \cH}{\partial q_j} - p_j \frac{\partial \cH}{\partial s} \\
    \dot s = p_i \frac{\partial \cH}{\partial p_i} - \cH.
  \end{cases}
\end{equation}
In what follows we will denote the {contact Hamiltonian} flow of $X_\cH$, or equivalently the solution of the above system, by
\begin{equation}
  \phi^\cH_{x_0} (t) := e^{t X_\cH} (x_0),
\end{equation} 
where $x_0\in M$ is the initial condition.

Differently from its symplectic counterpart, a contact Hamiltonian function is not conserved along the flow of $X_{\cH}$. Instead, its dynamics depends on $\reeb(\cH)$ {as can be noted in the following}
\begin{Prop}\rm
  The time derivative of $\cH$ along the flow of $X_\cH$ is
  \begin{equation}
    \dot{\cH} = -  \reeb(\cH)\cH.
  \end{equation}
\end{Prop}
This means, in particular, that the surface defined by $\cH=0$ is 
invariant under the contact Hamiltonian evolution and that this is generally not true for the other 
level sets~\cite{Bravetti2017a, Bravetti2017b, Liu_2021}.

Classical Hamiltonian systems can be studied in terms of the algebra of Hamiltonian vector fields on a symplectic manifold and in terms of the algebra of Hamiltonian functions with the Poisson bracket. This duality persists also in contact Hamiltonian mechanics.
The set of contact Hamiltonian functions on a contact manifold $(M,\eta)$ has a structure of a (local) Lie algebra in the sense of Kirillov~\cite{Kirillov1976}:
the bracket can be defined by \cite{Bravetti2020}
\begin{equation}
  \label{eq:jacobibrackets}
  \{f,g\}_\eta := - \eta \left([X_f,X_g]\right),
\end{equation}
which, in local canonical coordinates, has the form
\begin{equation}
  \left\{ f, g\right\}_\eta = \left(f \frac{\partial g}{\partial s} - g \frac{\partial f}{\partial s}\right)
  + p_j \left(\frac{\partial f}{\partial s} \frac{\partial g}{\partial p_j} - \frac{\partial g}{\partial s} \frac{\partial f}{ \partial p_j} \right) + \left(\frac{\partial f}{\partial q_j} \frac{\partial g}{\partial p_j} - \frac{\partial g}{\partial q_j} \frac{\partial f}{\partial p_j} \right).
  \label{eq:jacobibracketscoord}
\end{equation}
This bracket is bilinear, antisymmetric and satisfies the Jacobi identity, but not the Leibniz rule. Instead, one has
\begin{equation}
  \{f, g h\}_\eta = g \{f,h\}_\eta + h \{f,g\}_\eta + g h \{f ,1\}_\eta\,.
\end{equation}

For our purposes, a fundamental difference between the symplectic and the
contact settings is the fact that in the contact case
the map $a_{\eta}$, which assigns to each function 
the corresponding Hamiltonian vector field (see Definition~\ref{def:contactgradient}), is
 an isomorphism of Lie algebras, while in the symplectic case it is known to be just a homomorphism, as constant functions all belong to the kernel~\cite{de_Le_n_2019}.
This is the content of the next result~\cite{Albert_1989, de_Le_n_2019}.

\begin{theorem}\rm
  \label{thm:isomorphism}
  Given a contact manifold $(M,\eta)$,
  the map
  \begin{equation}
    a_\eta : \left(\mathcal{C}^\infty(M,\mathbb{R}),\{\cdot,\cdot\}_\eta\right) \to \left(\mathfrak{X}^C(M),[\cdot,\cdot]\right),
  \end{equation}
  is an isomorphism between the algebra of functions on $M$
  with the Jacobi bracket
  and the algebra $\mathfrak{X}^C(M)$ of infinitesimal contactomorphisms with the standard Lie bracket.
  Its inverse is provided by the contraction with the contact form:
  \(
    \eta : \left(\mathfrak{X}^C(M) , [\cdot,\cdot] \right) \to \left(\mathcal{C}^\infty(M,\mathbb{R}),\{\cdot,\cdot\}_\eta\right).
  \)
\end{theorem}

\begin{remark}\label{app:symplecticreview}\rm
  We can compare this with the usual Hamiltonian structures on symplectic manifolds.
A \emph{symplectic manifold} is a pair $(M,\omega)$, where $M$ 
is an even-dimensional manifold
and $\omega$ is a closed, non-degenerate,  $2$-form on $M$ called the \emph{symplectic form}~\cite{P.Libermann1987, Fasano2002}.
The symplectic form allows to associate to each function $\cH : M \to \mathbb{R}$ a vector field $X_\cH$, called the \emph{Hamiltonian vector field} or the \emph{symplectic gradient}, by
\begin{equation}
  \omega(X_\cH, \cdot) = - d\cH.
\end{equation}
In canonical coordinates $(q_i,p_i)$, the symplectic form is
\(
  \omega = d p_i \wedge d q_i
\)
and thus the integral curves of $X_\cH$ give
Hamilton's equations in their standard form
\begin{equation}\label{eq:SHameqs}
  \begin{cases}
    \dot{q} = \frac{\partial \cH}{\partial p} \\
    \dot{p} = - \frac{\partial \cH}{\partial q}\,.
  \end{cases}
\end{equation}
Finally, the \emph{Poisson bracket}~\cite{P.Libermann1987} of two functions is defined as
\begin{align} 
    \PB{f}{g} := \omega(X_g, X_f).
\end{align}
The side-to-side comparison of the symplectic and contact structures is summarized in Table~\ref{tab:comparison}.
\end{remark}

\begin{table}[h]
    \centering
    \begin{tabular}{lll}
    \toprule
    & Symplectic Hamiltonian Systems & Contact Hamiltonian Systems \\
    \midrule
    \midrule
     \multirow{2}{*}{Space} &
     \multicolumn{1}{p{6cm}}{$(M, \omega)$, $\dim M = 2n$,\newline $\omega$ closed 2-form s.t. $\omega^n\neq0$} &
     \multicolumn{1}{p{6.25cm}}{$(M, \eta)$, $\dim M = 2n+1$, \newline $\eta$ 1-form s.t. $\eta \wedge (d\eta)^n \neq 0$}
    \\
    \midrule
     \multicolumn{1}{p{2.3cm}}{Hamiltonian\newline vector fields} &
     \multirow{2}{*}{$\omega(X_\cH, \cdot) = - d\cH$} &
     \multicolumn{1}{p{6.25cm}}{$\eta (X_\cH) = - \cH$,\newline
      $d\eta(X_\cH, \cdot) = d\cH - d\cH(\reeb) \eta$}
    \\
     \multicolumn{1}{p{2.3cm}}{Algebra\newline on $C^\infty(M)$} & 
     \multirow{2}{*}{$\PB{f}{g} := \omega(X_g, X_f)$} &
     \multirow{2}{*}{$\{f,g\}_\eta := - \eta \left([X_f,X_g]\right)$}
    \\
    \multirow{2}{*}{Relation} & 
     \multicolumn{1}{p{6cm}}{$\left(\mathcal{C}^\infty(M),\{,\}_{\mathrm{PB}}\right) \to \left(\mathfrak{X}^H(M),[,]\right)$\newline homomorphism} & 
     \multicolumn{1}{p{6.25cm}}{$\left(\mathcal{C}^\infty(M),\{,\}_\eta\right) \to \left(\mathfrak{X}^C(M),[,]\right)$\newline isomorphism}
     \\
    \midrule
    \multicolumn{1}{p{2.3cm}}{Canonical\newline coordinates} &
     \multirow{2}{*}{$(q,p)$, $\omega = dp_i\wedge dq_i$} & 
     \multirow{2}{*}{$(q,p,s)$, $\eta = ds - p_i dq_i$, $\reeb=\frac{\partial}{\partial s}$}
    \\
    \multicolumn{1}{p{2.3cm}}{Equations of motion} &
     \multirow{2}{*}{$\dot{q} = \frac{\partial \cH}{\partial p}$,\ $\dot{p} = - \frac{\partial \cH}{\partial q}$} & 
     \multicolumn{1}{p{6.25cm}}{$\dot q = \frac{\partial \cH}{\partial p}$,\ $\dot p = -\frac{\partial \cH}{\partial q} - p\frac{\partial \cH}{\partial s}$,\newline $\dot s = p_i \frac{\partial \cH}{\partial p_i} - \cH$}
    \\
    \bottomrule
    \end{tabular}
    \caption{Comparison of symplectic and contact Hamiltonian structures.}
    \label{tab:comparison}
\end{table}

\section{The flow method}\label{sec:flow}

In this section we present the main idea of this work, proceeding in order of decreasing generality and increasing applicability.

Given a Lie algebra $\mathfrak{g}$ and two elements $A,B\in \mathfrak{g}$, in the following definition we outline a method to compute~\eqref{eq:logXY} without recurring to the general BCH formula~\eqref{eq:integralBCH}.
\begin{Def}\label{def:flowmethod}\rm
  Given a Lie algebra $\mathfrak{g}$, the {general} \emph{flow method} consists of the following: 
  \begin{enumerate}
    \item Find a Lie algebra isomorphism $\varphi$ between $\mathfrak{g}$ and some subalgebra
          $\mathfrak{g}_M$ of $\mathfrak{X}(M)$
          for some manifold $M$;
    \item Given $A,B\in\mathfrak{g}$, replace the product $e^A e^B$ by the composition of the flows
          $e^{t X_A}\circ e^{t X_B}$, where $X_A=\varphi(A)$ and $X_B=\varphi(B)$;
    \item Find the infinitesimal generator $Z(X_A,X_B)$ of $e^{t X_A}\circ e^{t X_B}$ {at $t=1$};
    \item Use the inverse isomorphism $\varphi^{-1}$ to find $Z(A,B)=\varphi^{-1}(Z(X_A,X_B))$.
  \end{enumerate}
\end{Def}

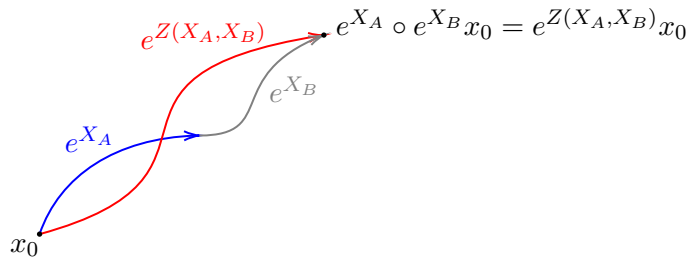
\begin{figure}[h!]
  \centering
    \tikzset{every picture/.style={line width=0.75pt}} 
    
    \begin{tikzpicture}[x=0.75pt,y=0.75pt,yscale=-1,xscale=1]
        
        \draw [color={rgb, 255:red, 0; green, 0; blue, 255 }  ,draw opacity=1 ]   (24.2,126.8) .. controls (34.25,97.25) and (62.34,78.57) .. (102.17,77.25) ;
        \draw [shift={(104,77.2)}, rotate = 178.88] [color={rgb, 255:red, 0; green, 0; blue, 255 }  ,draw opacity=1 ][line width=0.75]    (7.65,-2.3) .. controls (4.86,-0.97) and (2.31,-0.21) .. (0,0) .. controls (2.31,0.21) and (4.86,0.98) .. (7.65,2.3)   ;
        \draw [color={rgb, 255:red, 255; green, 0; blue, 0 }  ,draw opacity=1 ]   (24.2,126.8) .. controls (132.66,96.95) and (35.78,47.7) .. (164.17,27) ;
        \draw [shift={(166.12,26.69)}, rotate = 171.12] [color={rgb, 255:red, 255; green, 0; blue, 0 }  ,draw opacity=1 ][line width=0.75]    (7.65,-2.3) .. controls (4.86,-0.97) and (2.31,-0.21) .. (0,0) .. controls (2.31,0.21) and (4.86,0.98) .. (7.65,2.3)   ;
        \draw  [draw opacity=0][fill={rgb, 255:red, 0; green, 0; blue, 0 }  ,fill opacity=1 ] (22.87,126.8) .. controls (22.87,126.07) and (23.47,125.47) .. (24.2,125.47) .. controls (24.93,125.47) and (25.52,126.07) .. (25.52,126.8) .. controls (25.52,127.53) and (24.93,128.12) .. (24.2,128.12) .. controls (23.47,128.12) and (22.87,127.53) .. (22.87,126.8) -- cycle ;
        \draw [color={rgb, 255:red, 128; green, 128; blue, 128 }  ,draw opacity=1 ]   (104,77.2) .. controls (143.6,77.2) and (115.38,47.4) .. (164.6,27.29) ;
        \draw [shift={(166.12,26.69)}, rotate = 158.6] [color={rgb, 255:red, 128; green, 128; blue, 128 }  ,draw opacity=1 ][line width=0.75]    (7.65,-2.3) .. controls (4.86,-0.97) and (2.31,-0.21) .. (0,0) .. controls (2.31,0.21) and (4.86,0.98) .. (7.65,2.3)   ;
        \draw  [draw opacity=0][fill={rgb, 255:red, 0; green, 0; blue, 0 }  ,fill opacity=1 ] (164.79,26.69) .. controls (164.79,25.95) and (165.39,25.36) .. (166.12,25.36) .. controls (166.85,25.36) and (167.44,25.95) .. (167.44,26.69) .. controls (167.44,27.42) and (166.85,28.01) .. (166.12,28.01) .. controls (165.39,28.01) and (164.79,27.42) .. (164.79,26.69) -- cycle ;
        
        \draw (35.3,71.33) node [anchor=north west][inner sep=0.75pt]  [font=\small,color={rgb, 255:red, 0; green, 0; blue, 255 }  ,opacity=1 ]  {$e^{X_{A}}$};
        \draw (137.3,46.03) node [anchor=north west][inner sep=0.75pt]  [font=\small,color={rgb, 255:red, 128; green, 128; blue, 128 }  ,opacity=1 ]  {$e^{X_{B}}$};
        \draw (73.1,18.9) node [anchor=north west][inner sep=0.75pt]  [font=\small,color={rgb, 255:red, 255; green, 0; blue, 0 }  ,opacity=1 ]  {$e^{Z( X_{A} ,X_{B})}$};
        \draw (8,127.9) node [anchor=north west][inner sep=0.75pt]  [font=\small]  {$x_{0}$};
        \draw (170.58,11.65) node [anchor=north west][inner sep=0.75pt]  [font=\small]  {$e^{X_{A}} \circ e^{X_{B}} x_{0} =e^{Z( X_{A} ,X_{B})} x_{0}$};
  \end{tikzpicture}

  \caption{Graphical representation of steps 2.~and 3.~of the flow method.}
  \label{fig:flowmethod}
\end{figure}

A graphical representation of the idea behind steps 2.~and 3.~of the flow method is depicted in Fig.~\ref{fig:flowmethod}.

\begin{remark}\rm
  Clearly the flow method, whenever applicable, is a way to compute~\eqref{eq:logXY} without recurring to the BCH formula.
  An advantage of the flow method is that in some cases computing the composition of the flows
  $e^{t X_A}\circ e^{t X_B}$ can be very
  easy, whereas computing the corresponding product $e^A e^B$ can be rather cumbersome.
  {More importantly,
  once we have the composed flow $e^{t X_A}\circ e^{t X_B}$, then it is a standard calculation to recover the infinitesimal generator $Z(X_A,X_B)$.
  Therefore, step 3.~yields a substantial simplification.
  However, the main difficulty in the process is to find $M$ and $\mathfrak{g}_M$ such that 
  (i) there exists an isomorphism between $\mathfrak{g}$ and $\mathfrak{g}_M$
  and (ii) the flows $e^{t X_A}$ and $e^{t X_B}$ can be explicitly found. In general this is not even possible.
  However, as we show below, in some special cases one can work out the flow method and thus obtain closed-form expressions for the corresponding BCH formula.}
\end{remark}
\begin{remark}\rm
  The flow method was already used in a similar fashion in~\cite{Donnelly2005} in
  the context of finding the modified Hamiltonian for different symplectic numerical methods for
  the harmonic oscillator, as we will see in Example~\ref{ex: HOsymp}. However, the usefulness of this method has not been established before, perhaps because, as we will show shortly, the symplectic context is not the most useful one for the general purpose at hand.
\end{remark}

To implement the flow method, it turns out to be particularly convenient to choose $M$ to be either a symplectic or a contact manifold and $\mathfrak{g}_M$ to be a subalgebra of the Hamiltonian vector fields on $M$, which we will denote by $\mathfrak{h}_M$.
To avoid restating this all the time, we will denote the flow method with this particular choice of manifold and subalgebra the \emph{Hamiltonian flow method}, overloading the word Hamiltonian to denote both the contact and symplectic choices.

\begin{remark}\label{rmk:ham-is-nice}\rm
  An obvious advantage of the Hamiltonian flow method is the fact that in this case we are provided with an additional Lie algebra homomorphism $\psi$ between $\mathfrak{h}_M$ and the corresponding Hamiltonian functions.
  It is much easier to specify the desired Lie algebras in terms of Hamiltonian functions rather than in terms of vector fields directly. For instance, the Heisenberg Lie algebra is often written as the algebra of linear functions
  on a symplectic manifold together with their Poisson bracket, but rarely expressed as the corresponding Lie algebra of vector
  fields (and there is a reason for it, since the two in this case are not isomorphic, see below).
\end{remark}

Now, let $\{H_1,\dots,H_n\}$ denote a basis for $\psi(\mathfrak{h}_M)$, where $\psi$ is 
the Lie algebra homomorphism mentioned in Remark~\ref{rmk:ham-is-nice} above. Then a general
element in $\psi(\mathfrak{h}_M)$ takes the form
$$H:=\sum_{i=1}^n h_i H_i, \qquad \text{with} \qquad h_i\in \mathbb{R}\,. $$
This means that, given 2 elements $H_A=\sum_{i=1}^n a_i H_i$ and $H_B=\sum_{i=1}^n b_i H_i$, in order to find
$H_{Z(A,B)}=\sum_{i=1}^n z_i H_i$, we need to find the coefficients $z_i$ as functions of $a_i$ and $b_i$.

At this point, we need to add the following assumption.

\begin{Ass}\label{ass:crucial}\rm
  Assume that
  \begin{enumerate}[i.]
    \item $\psi:\mathfrak{h}_M\rightarrow\psi(\mathfrak{h}_M)$ is an isomorphism, and
    \item the flow $e^{t X_H}$ of the general $H$ (with unspecified coefficients $h_i$) can be obtained by quadratures.
  \end{enumerate}
\end{Ass}

If (and only if) Assumption~\ref{ass:crucial} is satisfied, then
points 3.~and 4.~in the general flow method (Definition~\ref{def:flowmethod}) can be replaced by
\begin{enumerate}
  \item[\~{3}a.] equate $e^{t X_{Z(A,B)}}=e^{t X_A}\circ e^{t X_B}$ to obtain a system of algebraic equations relating
    the known coefficients $a_i$ and $b_i$ to the unknown coefficients $z_i$
  \item[\~{3}b.] solve the above system to find the coefficients $z_i$ as functions of $a_i$ and $b_i$, say $z_i(a,b)$,
    thus
    obtaining $H_{Z(A,B)}=\sum_{i=1}^n z_i(a,b)H_i$
  \item[\~{4}.] then $Z(A,B)=\varphi^{-1}\psi^{-1}(H_{Z(A,B)})$, where $\psi$ is the isomorphism in 
  Assumption~\ref{ass:crucial} and $\varphi$ the one from step 1. of the flow method.
\end{enumerate}

\begin{Def}\label{def:implementablemethod}\rm
  We call the method given by the steps 1., 2., \~{3}a., \~{3}b., \~{4}.
  the \emph{{algebraic} Hamiltonian flow method}, or {A}HFM.
  The symplectic approach to the  {A}HFM will be specified by SAHFM, while the contact approach by CAHFM.
\end{Def}
Let us see how the {A}HFM works in the example, taken from~\cite{Donnelly2005},
of finding the modified Hamiltonian for a 1st order splitting integrator
for a classical harmonic oscillator. The purpose of this example is to develop the intuition on the flow method. At the same time, it also provides a first practical application (in discriminating among different numerical integrators).

\begin{example}[Splitting integrators for the harmonic oscillator]\rm
\label{ex: HOsymp}
Let us consider the 1-dimensional harmonic oscillator, with
$M=\mathbb{R}^2$, $\omega=dp\wedge dq$, and 
Hamiltonian function
\begin{equation}
    H := T(p) + V(q)\,,
\end{equation}
where 
\begin{equation}
    T(p)=\frac{p^2}{2}, \quad V(q)= \frac{q^2}{2}\,.
\end{equation}
The symplectic Hamiltonian vector fields corresponding to these functions are
\begin{equation}
    X_H = X_T + X_V, \quad X_T=p \frac{\partial}{\partial q}, \quad X_V=- q \frac{\partial}{\partial p}\,.
\end{equation}
  From the theory of geometric numerical integration
(see Section~\ref{sec:splittingint} or~\cite{Yoshida1990, Donnelly2005, Bravetti2020}),
we know that
\begin{equation}\label{eq:ho:S1}
S_1(\tau):=\exp\left\{\tau X_T\right\} \exp\left\{\tau X_V\right\},
\end{equation}
is a (symplectic) numerical integrator of 1st order for the flow of $H$, meaning that,
for small values of $\tau$, the timestep, one has
$S_1(\tau)(x_0) = \exp\{\tau X_H\}(x_0)+\cO(\tau)$ for any initial condition $x_0$. 
We also know that, being symplectic, $S_1(\tau)$ is  Hamiltonian with respect to some other Hamiltonian function $\tilde H(q,p;\tau)$, called the \emph{modified Hamiltonian} (see also e.g.~\cite{leimkuhler2004simulating,hairer2006geometric} for more details).
Finally, to find $\tilde H(q,p;\tau)$ we need to use the BCH formula~\eqref{eq:integralBCH} with $A=X_T$ and $B=X_V$.
Instead, in the following we will apply the flow method. 

We begin by observing that the commutation relations of our functions of interest with respect to the Poisson bracket $\PB{\cdot}{\cdot}$, see Remark~\ref{app:symplecticreview}, are
\begin{equation}
    \PB{T(p)}{V(q)} = - q\,p , \quad \PB{qp}{V(q)} = - 2 V(q) , \quad \PB{qp}{T(p)}= 2 T(p),
\end{equation}
so that $\mathcal{F}=\{V(q),T(p), q p\}$ provides a set of generators of a Lie subalgebra of $\left(\mathcal{C}^\infty(M,\mathbb{R}),\PB{\cdot}{\cdot}\right)$. 
Thus we can argue that the modified Hamiltonian $\tilde H(q,p;\tau)$ must be (for any $\tau$) a linear combination of the elements of the basis of this Lie subalgebra, that is,
\begin{equation}
\label{eqn:hammodHO}
    \tilde H(q,p;\tau) = 
    a(\tau) q^2 + b(\tau) p^2 + c(\tau) q\,p.
\end{equation}

Now, to compute the unknown coefficients $a(\tau),\ b(\tau),\ c(\tau)$  we observe that, for a fixed timestep $\tau$, we can solve the differential equation
\begin{equation}
    \frac{d x(t)}{d t}= X_{ \tilde H(q,p;\tau)}(x(t)),\qquad x(t):=(q(t),p(t)),
\end{equation}
that is,
\begin{equation}
    \begin{cases}
      \dot{q}(t) = 2 b(\tau) p(t) + c(\tau) q(t) \\
      \dot{p}(t) = -2 a(\tau) q(t) - c(\tau) p(t)
    \end{cases}.
\end{equation}
The solution, evaluated at $t=\tau$, is
\begin{equation}\label{eq:ho:flow}
    \begin{cases}
      q(\tau) = q_0 \cos \left(\omega \tau \right)+ \frac{(2 b p_0+c q_0) \sin \left(\omega \tau\right)}{\omega}\\
      p(\tau) = p_0 \cos ( \omega \tau) - \frac{(2 a q_0+c p_0) \sin \left( \omega \tau\right)}{\omega}
    \end{cases}.
\end{equation}
The map $x_0=x(0) \mapsto x_{1}=x(\tau)$ is $S_1(\tau)$, the numerical integrator map. Here we have used $\omega = -c^2 + 4 a b $. We know from~\eqref{eq:ho:S1} that the same function is given by the composition of the flows of $X_T$ and $X_V$ at $t=\tau$, i.e.~the maps
\begin{equation}
    \exp\left\{\tau X_V\right\} = \begin{cases}
      q_{1} = q_0\\
      p_{1} = - q_0 \tau + p_0\\
    \end{cases}
    \text{and} \quad
    \exp\left\{\tau X_T\right\} = \begin{cases}
      q_{1} = p_0 \tau + q_0\\
      p_{1} = p_0\\
    \end{cases}\,.
\end{equation}
We can now compute $a(\tau), b(\tau), c(\tau)$ by comparing
\begin{equation}\label{eq:ho:comparing}
    \begin{cases}
      q(\tau) = q_0 \cos \left(\omega \tau \right)+ \frac{(2 b p_0+c q_0) \sin \left(\omega \tau\right)}{\omega}=(1-\tau^2) q_0 + p_0\tau
      \\
      p(\tau) = p_0 \cos ( \omega \tau) - \frac{(2 a q_0+c p_0) \sin \left( \omega \tau\right)}{\omega}=p_0-q_0\tau\,,
    \end{cases}
\end{equation}
which should hold for any arbitrary initial condition $(q_0,p_0)$.
Solving the algebraic equations~\eqref{eq:ho:comparing}, we obtain
\begin{equation}
    a(\tau) =  b(\tau) = \Omega(\tau), \quad
    c(\tau) = \frac{\tau \ \Omega(\tau)}{2}, \qquad
    \Omega (\tau) := \frac{2 \arccos \left(1-\frac{\tau ^2}{2}\right)}{\tau  \sqrt{4-\tau^2}}\,,
\end{equation}
which in turn give the modified Hamiltonian~\eqref{eqn:hammodHO}. We notice that knowledge of the modified Hamiltonians for different integrators can help identify the most suitable one for the particular problem, as we will explain in more detail in Section~\ref{sec:splittingint}.
\end{example}

The purpose of the next example is to show that the flow method really relies on the contact algebra and cannot, in general, be used just by employing symplectic Hamiltonians.

\begin{example}[The Heisenberg algebra: SAHFM vs CAHFM]\label{ex:heisenberg}\rm
  Consider the Heisenberg algebra $\mathfrak{H}$.
  It is well-known~\cite{AIHPA_1970__13_2_103_0} that it can be represented as the Lie algebra
  of linear functions on the 2-dimensional symplectic manifold $(\mathbb{R}^2, \omega = dq \wedge dp)$, that is,
  \begin{equation}
    \mathfrak{H}:= \left(\Span_\mathbb{R} \left\{q ,p, 1\right\}, \PB{\cdot}{\cdot}\right),
  \end{equation}
  where $\PB{\cdot}{\cdot}$ denotes the standard Poisson bracket.
  The only non-vanishing commutator is $\PB{p}{q}=1$, as it should be.
  However, in this case there is no isomorphism between $\mathfrak{H}$ and a subalgebra of $\mathfrak{h}_M$ of symplectic Hamiltonian vector fields, as required by Definition~\ref{def:flowmethod}.
  This is because the symplectic gradient has a non-trivial kernel, consisting of all the constant functions~\cite{Fasano2002}. 
  A direct way to see this is by looking at Hamilton's equations in the symplectic case (equations~\eqref{eq:SHameqs} in Remark~\ref{app:symplecticreview}).
  Clearly, two Hamiltonians differing by an additive constant generate the same system of equations, and thus the same vector field.
  Therefore, in this example (and in the general case), step \~{4}.~of the SAHFM is problematic.
  Notice that this problem does not appear if we use the CAHFM.
  This is because in general the contact gradient defines an isomorphism, as we have seen in Theorem~\ref{thm:isomorphism}.
  Also in this case, looking back at the contact Hamiltonian equations can help  clarify this statement. Indeed, one can verify from equations~\eqref{eq:CHameqs}
  that in this case two Hamiltonians differing by an additive constant generate different vector fields. 
\end{example}

\section{Closed-form expressions for the BCH formula for various algebras} \label{sec:examples}

In this section we show how the flow method can  provide explicit closed forms of the BCH formula in a number of examples, namely,
the classical Heisenberg algebra, the contact Heisenberg algebra introduced in~\cite{Bravetti2017}, the quadratic symplectic algebra, and the quadratic contact algebra.
Finally, we use the latter to compute the modified Hamiltonians of several possible 
1st order contact splitting integrators for the damped harmonic oscillator~\cite{Bravetti2020}.

\subsection{The Heisenberg algebra}\label{sec:clHA}

Consider the classical Heisenberg algebra introduced in Example~\ref{ex:heisenberg}.
This algebra can be identified with
\begin{equation}
  (\Span_\mathbb{R} \{q ,p, {1}\}, \{\cdot,\cdot\}_{\eta} ),
\end{equation}
on the contact manifold $(\mathbb{R}^3,\eta = ds - p dq)$.
This new identification allows us to use the CAHFM, as discussed in Example~\ref{ex:heisenberg}.
More precisely, a general contact Hamiltonian function in the classical Heisenberg algebra has the form
\begin{equation}
  h = a q + b p + z {1},
  \quad a,b,z\in \mathbb{R},
  \label{eq:classicalheisembergfunction}
\end{equation}
so that its associated contact Hamiltonian vector field is
\begin{equation}
  X_h = b \frac{\partial}{\partial q} - a \frac{\partial}{\partial p} + (- a q - z) \frac{\partial}{\partial s}\,.
\end{equation}
The flow of this vector field for an initial condition $(q_0,p_0,s_0)$ is explicitly computed as
\begin{equation}
  \begin{cases}
    q(t) = b t + q_0  \\
    p(t) = -a t + p_0 \\
    s(t) = -\frac{1}{2} a b t^2- (a q_0 + z)t+s_0
  \end{cases}.
\end{equation}

\begin{figure}[h!]
  \centering
  \includegraphics[width=0.9\linewidth]{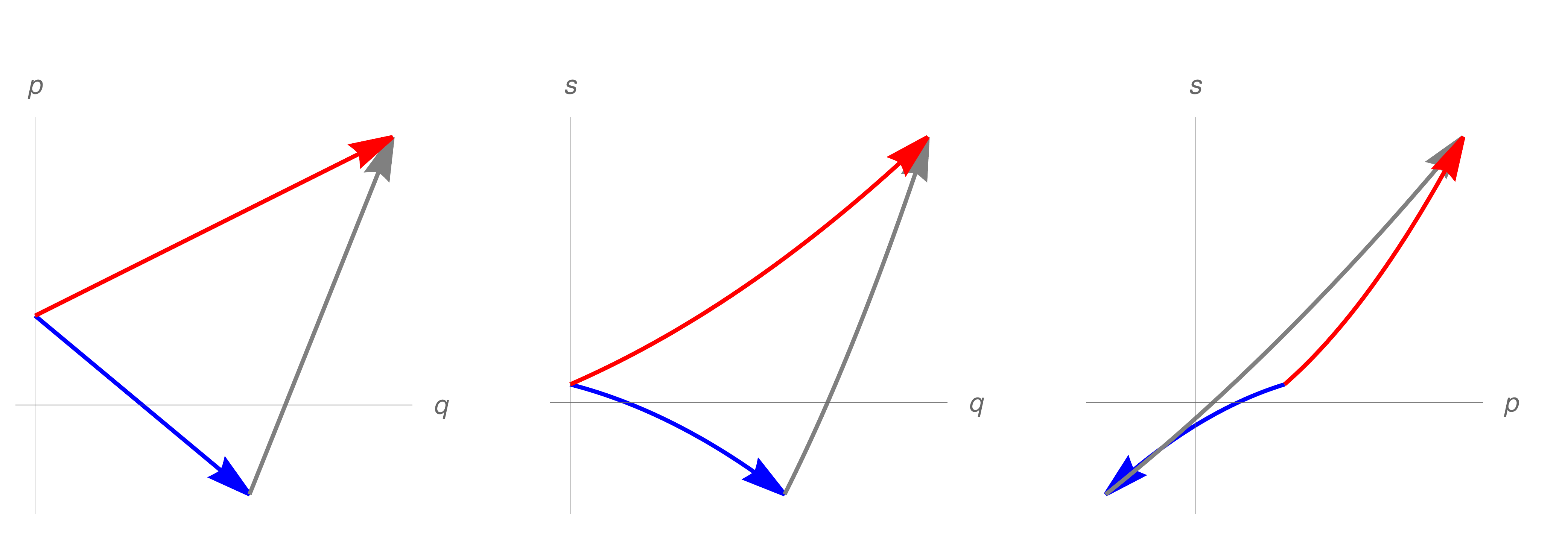}
  \caption{Graphical representation of the flow method applied to the Heisenberg algebra. The blue arrow is the flow of the first Hamiltonian $h$, the grey arrow is the flow of $\bar h$, and the red arrow is the flow of the Hamiltonian $\mathcal{h}$.}
  \label{fig:HAfig}
\end{figure}

Our aim is to compose the flows of two different contact Hamiltonians $h = a q + b p + z {1}$ and 
$\bar{h} = \bar{a} q + \bar{b} p + \bar{z} {1}$ and exploit the fact that such flow is a contact flow of some contact Hamiltonian $\mathcal{h}$ belonging to the same algebra.
Ultimately, this means that we are left to compute the explicit composition of the flows of $h$ and $\bar h$ and equate the coefficients with those of the flow of a general $\mathcal{h}=\alpha q + \beta p + \zeta {1}$  
in order to find the appropriate $\mathcal{h}$.
In practice, this amounts to solving a system of algebraic equations obtained as follows:
the composition of the flows of $h$ and $\bar h$ at some time $t$, namely ${e^{t X_{\bar{\mathcal{h}}}}} \circ {e^{t X_{\mathcal{h}}}}$, is computed explicitly as
\begin{equation}
  \begin{cases}
    q(t) = (b + \bar{b}) t + q_0  \\
    p(t) = -(a + \bar{a}) t + p_0 \\
    s(t) = -\frac{1}{2} t^2 (a b+\bar{a} (2 b+\bar{b}))-t(q_0 (a+\bar{a})+z+\bar{z})+s_0\,.
  \end{cases}.
\end{equation}
By {equating} this with the flow of $\mathcal h$, namely
\begin{equation}
  \begin{cases}
    q(t) = \beta t + q_0  \\
    p(t) = -\alpha t + p_0 \\
    s(t) = -\frac{1}{2} \alpha \beta t^2- (\alpha q_0 + z)t+s_0\,,
  \end{cases}
\end{equation}
at $t=1$, 
we obtain the following defining relations for the coefficients
\begin{equation}
  \begin{cases}
    \alpha = a + \bar{a},                                           \\
    \beta = b + \bar{b},                                            \\
    \zeta = \frac{1}{2} (-a \bar{b} +\bar{a} b +2 z+2 \bar{z})\,, \\
  \end{cases}
\end{equation}
which indeed is the standard closed-form expression for the BCH formula for the Heisenberg algebra. 
In Figure~\ref{fig:HAfig} we give a graphical representation of the method.

\subsection{The contact Heisenberg algebra}\label{sec:coHA}

\begin{figure}[h!]
  \centering
  \includegraphics[width=0.9\linewidth]{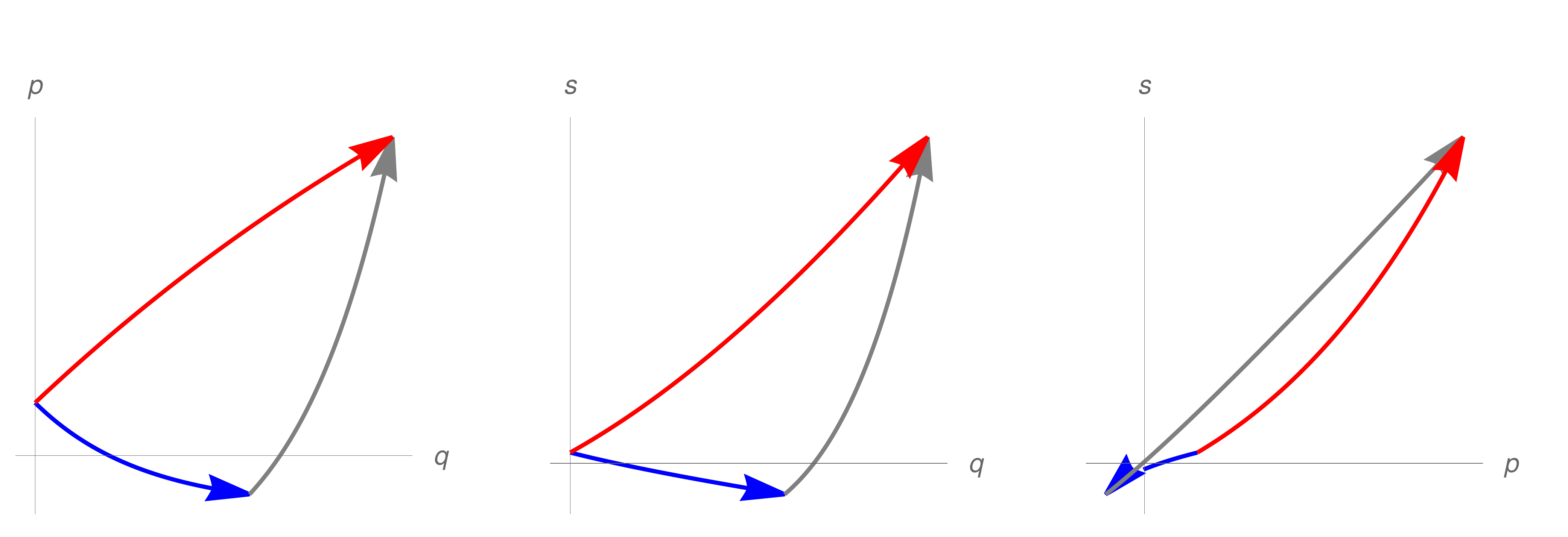}
  \caption{Graphical representation of the flow method applied to the contact Heisenberg algebra.
  The blue arrow is the flow of the first Hamiltonian $f$, the grey arrow is the flow of $\bar f$, and the red arrow is the flow of the Hamiltonian $\mathcal{f}$.}
  \label{fig:CHAfig}
\end{figure}

The contact Heisenberg algebra (CHA) is a natural extension of the classical Heisenberg algebra \cite{Bravetti2017} in the contact phase space $(\mathbb{R}^3,\eta= ds - p dq)$, being defined as
\begin{equation}
  (\Span_\mathbb{R} \{q ,p , s, {1}\}, \{\cdot,\cdot\}_{\eta} )\,.
\end{equation}
Therefore, any contact Hamiltonian function 
in this algebra has the following form:
\begin{equation}
  f = a q + b p + c s + z {1},
  \quad a,b,c,z\in\mathbb{R}.
\end{equation}
and the corresponding contact Hamiltonian vector field is
\begin{equation}
  X_f = b \frac{\partial}{\partial q} + (-a - p c) \frac{\partial}{\partial p} + ( - a q - c s - z) \frac{\partial}{\partial s}\,.
\end{equation}
Again, we can compute the flow explicitly as
\begin{equation}
  \begin{cases}
    q(t) = b t + q_0                                             \\
    p(t) = - \frac{a}{c} \left(1 - e^{-ct}\right) + e^{-c t} p_0 \\
    s(t) = s_0 e^{-ct} - q_0 \frac{a}{c}\left(1 - e^{-ct} \right) + \frac{a b}{c^2} - \frac{a b}{c} t - \frac{c}{z} + e^{-ct} \left( \frac{z}{c} - \frac{a b}{c^2} \right)\,,
  \end{cases}
\end{equation}
and thus consider the composition of the flows of two contact Hamiltonians
\begin{equation}
  f = a q + b p + c s + z {1} \qquad \text{and} \qquad
  \bar{f} = \bar{a} q + \bar{b} p + \bar{c} s + \bar{z} {1};
\end{equation}
and compare it with the flow generated by 
\begin{equation}
  \mathcal{f} = \alpha q + \beta p + \gamma s + \zeta {1}
\end{equation}
at $t=1$.
This leads to the following solution
\begin{equation}
  \begin{cases}
  \gamma =  c + \bar{c} \\
    \zeta =\left(\frac{\gamma e^{\gamma}}{e^{\gamma}-1}\right)
    \bigg(
    \left(e^{-c}-1\right) \left(\frac{e^{\bar{c}} (ab - cz)}{c^2} +\frac{(\bar{a} \bar{b} - \bar{c}\bar{z})}{\bar{c}^2} \right)   
    +\frac{e^{-\bar{c}} a b \bar{c} + \bar{a} b c\left(1-e^{-\bar{c}}\right) + \bar{a} \bar{b} c}{c \bar{c}}
    - \frac{\beta \alpha\left(e^{-\gamma} + \gamma + 1\right)}{\gamma^2}
    \bigg)
    \\
    \alpha = \frac{\gamma \left(a \bar{c} \left(e^{c }-1\right)+\bar{a} c e^{c} \left(e^{\bar{c} }-1\right)\right)}{c \bar{c} \left(e^  \gamma-1\right)} \\
    \beta = b + \bar{b}\\
  \end{cases},
  \label{eqn:solutioncontactheisemberg}
\end{equation}
which again is the correct closed-form expression for the BCH formula for this algebra (cf.~\cite{Bravetti2017}).
The method is represented graphically in Figure~\ref{fig:CHAfig}.

\subsection{The quadratic symplectic algebra} \label{SymplecQAlg}

The quadratic symplectic algebra is the algebra in the two-dimensional symplectic manifold $(\mathbb{R}^2,\ \omega = dp \wedge dq)$ generated by the three quadratic functions, that is, 
\begin{equation}
\label{eqn:qsympalg}
  \left( \Span_{\mathbb{R}}\{p^2, q^2, q p\}, \{\cdot,\cdot\}_{(PB)} \right).
\end{equation}
Notice that this case includes Example~\ref{ex: HOsymp} and that here we do not have constant functions as generators of the algebra, differently from Example~\ref{ex:heisenberg}.
Therefore, there is an isomorphism between this Lie algebra and the algebra of the corresponding symplectic Hamiltonian vector fields. 
Indeed, to each $f(q,p) = a q^2 + b p^2 + c q p$, $a,b,c\in\mathbb{R}$, we can associate uniquely 
\begin{equation}
    X_f = (c q + 2 b p) \frac{\partial}{\partial p} + (-c p - 2 a q) \frac{\partial}{\partial q}\,.
\end{equation}

Hamilton's equations for $X_f$ are given by the linear ordinary differential equation
\begin{equation}
  \label{eqn:linearsymplectic}
  \begin{pmatrix}
    \dot{q} \\ \dot{p}
  \end{pmatrix} = \begin{pmatrix}
    c & 2 b \\ -2 a & -c
  \end{pmatrix}
  \begin{pmatrix}
    q \\ p
  \end{pmatrix},
\end{equation}
whose general solution at $t=1$ for an initial condition $x(0)=(q_0,p_0)$ can be expressed as
\begin{equation}
    \begin{pmatrix}
      q(1) \\ p(1)
    \end{pmatrix}
    = \left( \cosh(\det(A)) \mathbb{I} + \frac{\sinh(\det(A))}{\sqrt{\det{(A)}}} A \right) \begin{pmatrix}
    q_0 \\ p_0  
    \end{pmatrix}.
\end{equation}
Now we want to compose two of these flows, generated respectively by $H_A = a_1 q^2 + b_1 p^2 + c_1 q p$ and $H_B=a_2 q^2 + b_2 p^2 + c_2 q p$, which correspond to the flows given by \eqref{eqn:linearsymplectic} with the matrices
\begin{equation}
    A=\begin{pmatrix}
      c_1 & 2 b_1 \\ -2a_1 & -c_1
    \end{pmatrix}\qquad \text{and} \qquad B=\begin{pmatrix}
      c_2 & 2 b_2 \\ -2a_2 & -c_2
    \end{pmatrix}.
\end{equation}
To keep the following expressions compact, it is convenient to define the notations
\begin{align}
    rd(A) := \sqrt{\det(A)}\,, 
    \quad
    sd(A) := \sinh(\det{A})
    \quad\mbox{and}\quad
    cd(A) := \cosh(\det{A})\,.
\end{align}
Then the composition of the two different flows given by these two Hamiltonian matrices~\cite[Chapter X]{Fasano2002} can be expressed as
\begin{align}
\label{eqn:firstcompSQA}
    &\left( cd(A) \mathbb{I} + \frac{sd(A)}{rd(A)}  A \right) \left( cd(B) \mathbb{I} + \frac{sd(B)}{rd(B)}  B \right) = \\
    &\qquad= cd(A) cd(B) \mathbb{I} + cd(A) \frac{sd(B)}{ rd(B)}  B 
    + cd(B) \frac{sd(A)}{rd(A)}  A + \frac{sd(A)}{rd(A)} \frac{sd(B)}{rd(B)} A B.
\end{align}
Moreover, since $A$ and $B$ are Hamiltonian matrices, their product can be decomposed as follows
\begin{equation}
    A B = J S + \lambda \mathbb{I}\,,
\end{equation}
where $J=\begin{pmatrix}
  0 & 1\\ -1 & 0
\end{pmatrix}$, $\lambda=-2 a_1 b_2-2 a_2 b_1+c_1 c_2$,  
and $S$ is the symmetric matrix
\begin{equation}
    S=\begin{pmatrix}
 \frac{1}{2} (4 a_1 c_2 -4 a_2 c_1) & \frac{1}{2} (4 a_1 b_2-4 a_2 b_1) \\
 \frac{1}{2} (4 a_1 b_2-4 a_2 b_1) & \frac{1}{2} (4 b_2 c_1-4 b_1 c_2) 
    \end{pmatrix}\,.
\end{equation}
Substituting, and exploiting $J^{-1}=-J$, we can rewrite the right-hand side of~\eqref{eqn:firstcompSQA} as
\begin{align}
    &\left(cd(A) cd(B) + \frac{sd(A)}{rd(A)} \frac{sd(B)}{rd(B)} \lambda \right) \mathbb{I} +
    \frac{sd(A)}{rd(A)} \frac{sd(B)}{rd(B)} J S 
    +cd(A) \frac{sd(B)}{rd(B)}  B + cd(B) \frac{sd(A)}{rd(A)}  A. 
\end{align}
Comparing this result with the flow generated by a third matrix $C$ of the same type, we obtain the two relations
\begin{equation}
\label{eqn:systemsymp}
    \begin{cases}
       \frac{sd(A)}{rd(A)} \frac{sd(B)}{rd(B)}  S - cd(A) \frac{sd(B)}{rd(B)} J B - cd(B) \frac{sd(A)}{ rd(A)} J A =  - \frac{sd(C)}{rd(C)} J C \\
       cd(A) cd(B) + \frac{sd(A)}{rd(A)} \frac{sd(B)}{rd(B)} \lambda = cd(C)\,.
    \end{cases}
\end{equation}
Form the second equation we can obtain immediately the determinant of $C$. Moreover, using the inverse function of the hyperbolic cosine, we can also obtain the corresponding value of the hyperbolic sine by
\begin{equation}
    \sqrt{\det(C)} = \ln\left(x + \sqrt{x^2-1}\right) \Rightarrow sd(C) = \sqrt{x^2+1}\,,
\end{equation}
where
\begin{equation}
\label{eqn:xSQA}
    x = cd(A) cd(B) + \frac{sd(A)}{rd(A)} \frac{sd(B)}{rd(B)} \lambda\,.
\end{equation}
Now, using the second equation in~\eqref{eqn:systemsymp}, we obtain the final result
\begin{align}
    \label{eqn:finalqsa}
    C=&\,\frac{\ln(x+\sqrt{x^2-1})}{\sqrt{x^2-1}} \bigg(\frac{sd(A)}{rd(A)} \frac{sd(B)}{rd(B)}  S 
    - cd(A) \frac{sd(B)}{rd(B)} J B - cd(B) \frac{sd(A)}{rd(A)} J A \bigg)\,,
\end{align}
which is the closed-form expression for the BCH formula for the quadratic symplectic algebra (in matrix form).
To our knowledge this result has not been reported elsewhere previously and therefore it is interesting in itself.

\begin{figure}[h!]
  \centering
  \includegraphics[width=0.9\linewidth]{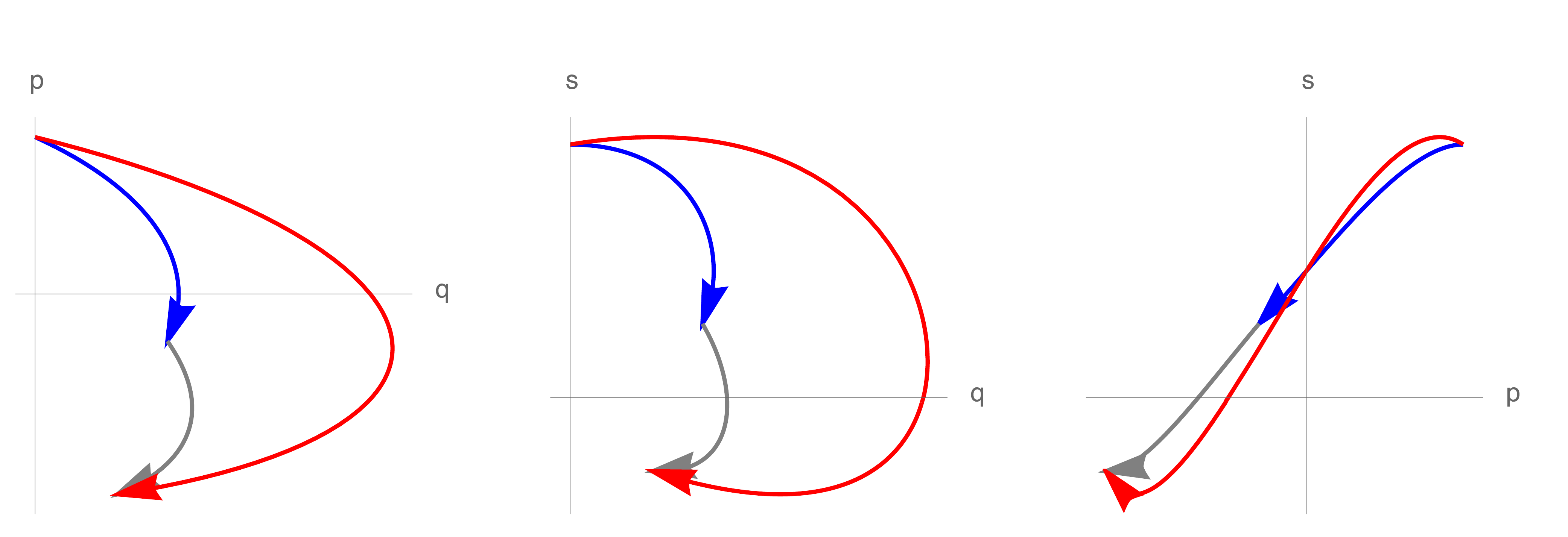}
  \caption{Graphical representation of the flow method applied to the quadratic symplectic algebra.
   The blue arrow is the flow of the first Hamiltonian $H_A$, the grey arrow is the flow of $H_B$, and the red arrow is the flow of the Hamiltonian $H_C$.}
  \label{fig:sympfig}
\end{figure}

The graphical result of this procedure is plotted in Figure~\ref{fig:sympfig}. 
Since we have been working on the two-dimensional manifold $(\mathbb{R}^2,\omega=dp \wedge dq)$, one may be surprised by the appearance of the $(q,s)$ and $(p,s)$ planes in Figure~\ref{fig:sympfig}. Indeed, we are describing there the flow method applied to the extension of the same algebra to the contact manifold $(\mathbb{R}^2\times \mathbb{R}, \eta = ds - p dq)$. In practice, the $(p,q)$-projection (leftmost plot) is exactly what we have computed here, and for the rest of the pictures we only need to extend~\eqref{eqn:linearsymplectic} with the additional differential equation $\dot{s} = b p^2 - a q^2$.

The relevance of this example will be clear in the next section on the quadratic contact algebra. There we will need to use a combination of the ideas presented here and in the previous examples.

\subsection{The quadratic contact algebra} \label{sec:QCA}

Another interesting subalgebra of the contact Hamiltonian algebra is the quadratic contact algebra (QCA), defined on the three dimensional contact manifold $(\mathbb{R}^3, \eta=ds-pdq)$ as
\begin{equation}\label{eq:QCA}
  (\Span_\mathbb{R} \{p^2 ,q^2, q p, s, {1} \}, \{ \cdot, \cdot\}_\eta).
\end{equation}
This algebra is interesting since it contains the quadratic symplectic algebra, the conformal symplectic one (generated by $\{q^2,p^2,q p ,s\}$), and the Hamiltonian function of the harmonic oscillator (with and without dissipation).
We will come back to the latter in the next subsection, where we will rely on the results of this section to compute the modified Hamiltonian for different choices of splitting numerical integrators for such system.

Hamiltonian functions in the algebra~\eqref{eq:QCA} 
have the form
\begin{equation}
  f = a q^2 + b p^2 + c s + d q p + z {1},
  \quad a, b, c, d, z \in\mathbb{R}\,,
  \label{eq:generalfuncitonquadratic}
\end{equation}
and their corresponding contact gradients are
\begin{equation}
  \begin{cases}
    \dot{q} = \frac{\partial f}{\partial p} = 2 b p + d q                                             \\
    \dot{p} = - \frac{\partial f}{\partial q} - p \frac{\partial f}{\partial s} = - 2 a q - (d + c) p \\
    \dot{s} = p\frac{\partial f}{\partial p} - f=b p^2 - a q^2 - c s - z\,.
  \end{cases}
\end{equation}
This system of differential equations is integrable by quadratures, 
since the first two equations are linear and the third one is linear in $s$ once the first two are solved.
In the same way as in the previous sections, we compose the flows of two contact Hamiltonians in the algebra,
\begin{equation}
  f = a q^2 + b p^2 + d s + c q p + z {1} \quad \text{and} \quad \bar{f} = \bar{a} q^2 + \bar{b} p^2 + \bar{d} s + \bar{c} q p + \bar{z} {1},
\end{equation}
and compare the result with the flow of a third Hamiltonian
\begin{equation}
  \mathcal{f} = \alpha q^2 + \beta p^2 + \gamma q p + \delta s + \zeta {1}.
\end{equation}
The flow method immediately provides an expression for the unknowns $\delta$ and $\zeta$ from the solution $s(t)$, reading
\begin{equation}
\label{eqns:deltazetacontquad}
  \begin{cases}
    \delta = d + \bar{d} \\
    \zeta = \frac{\delta (- \bar{d} z + \bar{d} e^d z - d e^d \bar{z} + d e^\delta \bar{z} )}{d \bar{d} (-1+e^\delta)}\,.
  \end{cases}
\end{equation}

To obtain the remaining parameters $\alpha$, $\beta$ and $\gamma$ we repeat the same computation as in the previous section: observe that the solution in the plane $(q,p)$ is independent of the dynamics in $s$, so we only need to solve the dynamical system
\begin{equation}
\label{eqn:diffeqn}
  \begin{pmatrix}
    \dot{q} \\ \dot{p}
  \end{pmatrix} =
  \underbrace{\begin{pmatrix}
    c    & 2 b    \\
    -2 a & -c - d
  \end{pmatrix}}_{A}
  \begin{pmatrix}
    q \\ p
  \end{pmatrix}.
\end{equation}
The solution of equation \eqref{eqn:diffeqn} at $t=1$ can be expressed as
\begin{equation}
\label{eqn:solCQA}
\begin{pmatrix}
  q(1) \\p(1)
\end{pmatrix} = e^{-\frac{d}{2}} \left( cd(\Tilde{A}) \mathbb{I} + \frac{sd(\Tilde{A})}{\det(\Tilde{A})} \Tilde{A} \right) \begin{pmatrix}
  q(0) \\ p(0)
\end{pmatrix},
\end{equation}
where $\Tilde{A}$ is the matrix
\begin{equation}
\label{eqn:tildaCHA}
    \Tilde{A} = \begin{pmatrix}
        c +d   & 2 b    \\
    -2 a & -c - d
    \end{pmatrix},
\end{equation}
which is a Hamiltonian matrix, as in the previous section.

\begin{figure}[h!]
  \centering
  \includegraphics[width=0.9\linewidth]{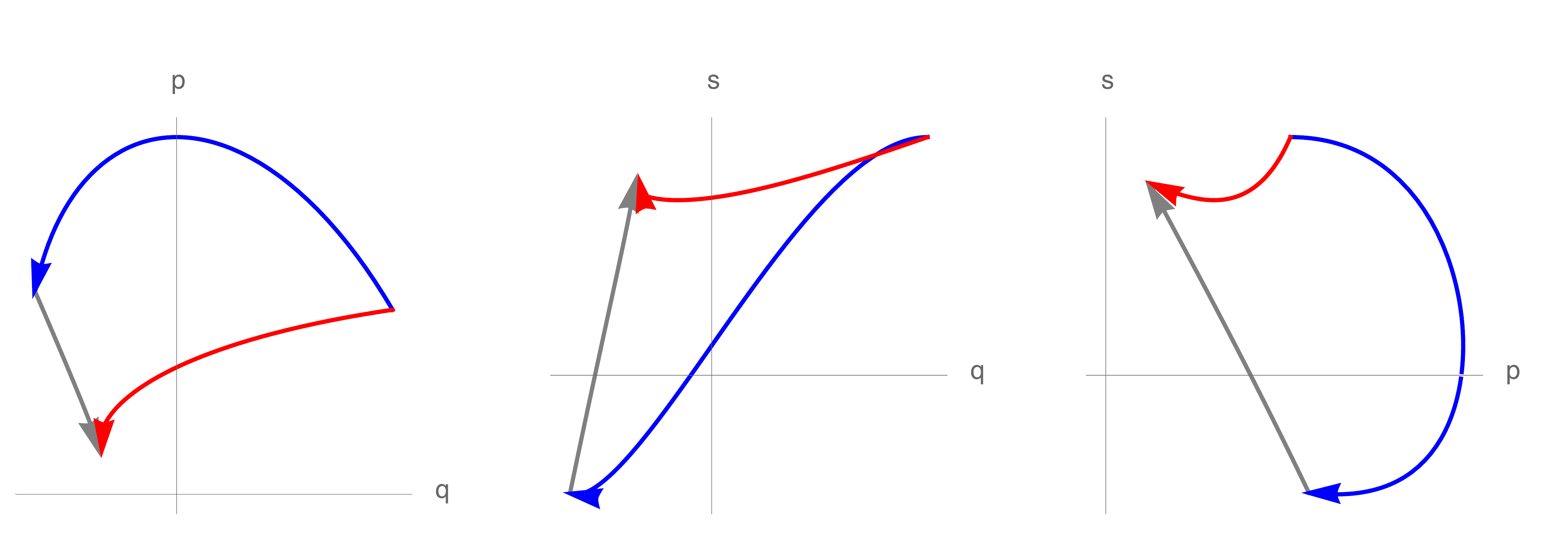}
  \caption{Graphical representation of the flow method applied to the quadratic contact algebra.}
  \label{fig:contquad}
\end{figure}

When we compose the flows generated by the matrix $A$ in~\eqref{eqn:diffeqn} and the analogous matrix
$B=\begin{pmatrix}
  \bar{c} & 2 \bar{b} \\ - 2 \bar{a} & -\bar{c} -\bar{d}
 \end{pmatrix}$ at $t=1$, we obtain
\begin{align}
    e^{-\frac{d+\bar{d}}{2}} \left( cd(\Tilde{A}) \mathbb{I} + \frac{sd(\Tilde{A})}{\det(\Tilde{A})} \Tilde{A} \right)  \left( cd(\Tilde{B}) \mathbb{I} + \frac{sd(\Tilde{B})}{\det(\Tilde{B})} \Tilde{B} \right).
\end{align}
This has to be compared with the flow generated by
\begin{equation}
    \Tilde{C}=\begin{pmatrix}
      \gamma & 2 \beta \\
      - 2 \alpha & - \gamma - \delta
    \end{pmatrix},
\end{equation}
which, at $t=1$, has the same form as equation~\eqref{eqn:solCQA}. 
Since we already know the equation for the parameter $\delta$ from~\eqref{eqns:deltazetacontquad}, we can reduce this condition to~\eqref{eqn:systemsymp} from the symplectic case: the exponential term $\exp(d/2 + \bar{d}/2)$ simplifies and the remaining terms are the flows generated by Hamiltonian matrices of the form~\eqref{eqn:tildaCHA}.
This leads to the solution for $\Tilde{C}$ being given by
    \begin{align}\label{eq:tildeC}
    \Tilde{C}=&\,\frac{\ln(x+\sqrt{x^2-1})}{\sqrt{x^2-1}} \bigg(\frac{sd(\Tilde{A})}{rd(\Tilde A)} \frac{sd(\Tilde{B})}{rd(\Tilde B)} \Tilde{S}  
    - cd(\Tilde{A}) \frac{sd(\Tilde{B})}{rd(\Tilde B)} J \Tilde{B} - 
    cd(\Tilde{B})\frac{sd(\Tilde{A})}{rd(\Tilde A)} J \Tilde{A} \bigg)
    \end{align}
where $x$ is defined as in \eqref{eqn:xSQA} using $\Tilde{A}$ and $\Tilde{B}$.
Once again this equation gives the closed-form expression for the BCH formula for this algebra.
{Note that for $z=\bar z=0$ we recover a closed-form expression for the conformal symplectic algebra.
To our knowledge this result has not been reported elsewhere
previously and therefore it is interesting in itself}.
A graphical test of the results is shown in Figure~\ref{fig:contquad}.

\subsection{The complexified $su(2)$}

 In this section we will show how to map the complexification of $su(2)$, $su(2)_{\mathbb{C}} \simeq sl(2, \mathbb{C}) \simeq su(2) \otimes \mathbb{C}$, to a complexified version of the quadratic symplectic Lie algebra of Section~\ref{SymplecQAlg}. To keep the presentation short we will omit the details of the complexification procedure, for which we refer the readers to \cite[Chapters 3.6 and 4.6]{Hall2015}.

The Lie algebra $su(2)$ can be represented as the vector space
\begin{equation}
    \Span_{\mathbb{R}} \left\{ \Sigma_1 := \imath \sigma_1, \Sigma_2 := -\imath \sigma_2, \Sigma_3 := \imath \sigma_3 \right\},
\end{equation}
spanned by the Pauli matrices $\sigma_1, \sigma_2, \sigma_3$.
The structure coefficients of the algebra are well known, and with our notation they correspond to
\begin{equation}
    \left[\Sigma_k, \Sigma_l\right] = 2 \epsilon_{k\, l\, m} \Sigma_m, \quad k,l,m = 1, \ldots, 3.
\end{equation}

The complexification of $su(2)$ leads to the Lie algebra $su(2)_{\mathbb{C}} \simeq su(2)\otimes \mathbb{C}$ spanned by
\begin{equation}
    H = - \imath \Sigma_3, \quad X = - \frac{\imath}{2} (\Sigma_1 - \imath \Sigma_2),\quad Y = - \frac{\imath}{2} (\Sigma_1 + \imath \Sigma_2).
\end{equation}
These new generators satisfy the commutation relations \cite[Chapter 4.6]{Hall2015}:
\begin{equation}
    \left[H,X\right] = 2 X, \quad \left[H,Y\right] = -2Y,\quad \left[X,Y \right] = H.
\end{equation}

To find a mapping between the $su(2)$ and the symplectic algebra, we start from the commutation relation
\begin{equation}
    \{q^2, p^2\}_{PB} = 4 q p, \quad \{p^2, q p\}_{PB} = -2 p^2, \quad \{q p, q^2\}_{PB} = - 2 q^2.
\end{equation}
It is a matter of explicit computation to find the mapping
\begin{equation}
   \{q^2, p^2, qp\} \mapsto \{ \imath \Sigma_1 - \Sigma_2,  -\imath \Sigma_1 - \Sigma_2, - \imath \Sigma_3\},
\end{equation}
and its inverse
\begin{equation}
    \{\Sigma_1, \Sigma_2, \Sigma_3\} \mapsto \left\{\frac{\imath}{2}(p^2 - q^2), -\frac{1}{2}(p^2 + q^2), \imath qp\right\}.
\end{equation}

With this mapping, an element of the $su(2)$ algebra is mapped to an element of the complexified symplectic quadratic algebra by:
\begin{equation}\label{iso-to-su2}
    \chi = \sum_{k=1}^{k=3} \mu_k \Sigma_k \quad\mapsto\quad \bar{\chi} = \underbrace{\left( - \frac{\imath}{2}\mu_1 - \frac{\mu_2}{2}\right)}_{a} q^2 + \underbrace{\left(\frac{\imath}{2} \mu_1 -\frac{\mu_2}{2}\right)}_{b} p^2 + \underbrace{\left( \imath \mu_3\right)}_{c} q p.
\end{equation}
To obtain the corresponding BCH formula, is then enough to consider the matrix
\begin{equation}
    {\bf A} = \begin{pmatrix}
        c & 2 b\\
        -2a & -c       
    \end{pmatrix}
    = \begin{pmatrix}
        \imath \mu_3 & \imath \mu_1 - \mu_2\\
        \imath \mu_1 + \mu_2 & -\imath \mu_3
    \end{pmatrix},
\end{equation}
and the respective matrix ${\bf B}$, in equation~\eqref{eqn:finalqsa} and then invert the mapping \eqref{iso-to-su2}.
This also provides an alternative derivation of the well-known parameterization of the composition law of the group $SU(2)$ in terms of Pauli matrices.

Now, using the second equation in~\eqref{eqn:systemsymp}, we obtain the final result
\begin{align}
    \label{eqn:finalsu2}
    {\bf C} =&\,\frac{\ln({\bf x} +\sqrt{{\bf x}^2-1})}{\sqrt{{\bf x}^2-1}} \bigg(\frac{sd({\bf A})}{rd({\bf A})} \frac{sd({\bf B})}{rd({\bf B})}  {\bf S} - cd({\bf A}) \frac{sd({\bf B})}{rd({\bf B})} J {\bf B} - cd({\bf B}) \frac{sd({\bf A})}{rd({\bf A})} J {\bf A} \bigg)\,,
\end{align}
where 
\begin{align}
{\bf x} &= cd({\bf A}) cd({\bf B}) + \frac{sd({\bf A})}{rd({\bf A})} \frac{sd({\bf B})}{rd({\bf B})} \Lambda\, , \\
{\bf S} & =\begin{pmatrix}
 \mu_1 \nu_3 - \mu_3 \nu_1 + (\mu_3 \nu_2 - \mu_2 \nu_3) \imath & (\mu_1 \nu_2 - \mu_2 \nu_1) \imath \\
 (\mu_1 \nu_2 - \mu_2 \nu_1) \imath & \mu_1 \nu_3 - \mu_3 \nu_1 + (\mu_2 \nu_3 - \mu_3 \nu_2) \imath 
    \end{pmatrix}\,.
\end{align}
and $\Lambda = - \mu_1 \nu_1 - \mu_2 \nu_2 - \mu_3 \nu_3$.

\subsection{Splitting numerical integrators} \label{sec:splittingint}

The flow method can immediately be put to use in the computation of the modified Hamiltonian of certain splitting integrators.
Contact splitting integrators are a family of numerical algorithms to integrate flows of Hamiltonian functions 
that can be written as sums of separate terms, i.e.,
\begin{equation}
    \cH = \sum_{i=1}^{n} \ch_i.
\end{equation}
It is convenient to assume that each $\ch_i$ is a Hamiltonian function whose flow can be explicitly integrated. 
Then, for a small timestep $0 < \tau \ll 1$, the flow $\phi^\cH_x(\tau)$ of $\cH$ can be approximated to first order in $\tau$ by the map
\begin{equation}
    \phi^\cH_x(\tau) = \left(\bigcirc_{i=1}^{n} \exp\{\tau X_{\ch_i}\}\right)(x) + \mathcal{O}(\tau),
\end{equation}
where $\bigcirc_{i=1}^{n} \psi_i := \psi_1 \circ \psi_2 \circ \cdots \circ \psi_n$,
and to second order by
\begin{equation}
    \phi^\cH_x(\tau) = \left(\bigcirc_{i=1}^{n-1} \exp\left\{\frac{\tau}{2} X_{\ch_i}\right\} \circ \exp\{\tau X_{\ch_n}\} \circ  \bigcirc_{i=n-1}^{1} \exp\left\{\frac{\tau}{2} X_{\ch_i}\right\}\right)(x) + \mathcal{O}(\tau^2).
\end{equation}
The \emph{modified Hamiltonian} of the numerical integrator is the Hamiltonian function that generates the integrator, 
that is, the Hamiltonian of the approximate flow (see e.g.~\cite{Bravetti2020} for more details).

\begin{figure}[h]
  \centering
  \includegraphics[width=1.\linewidth]{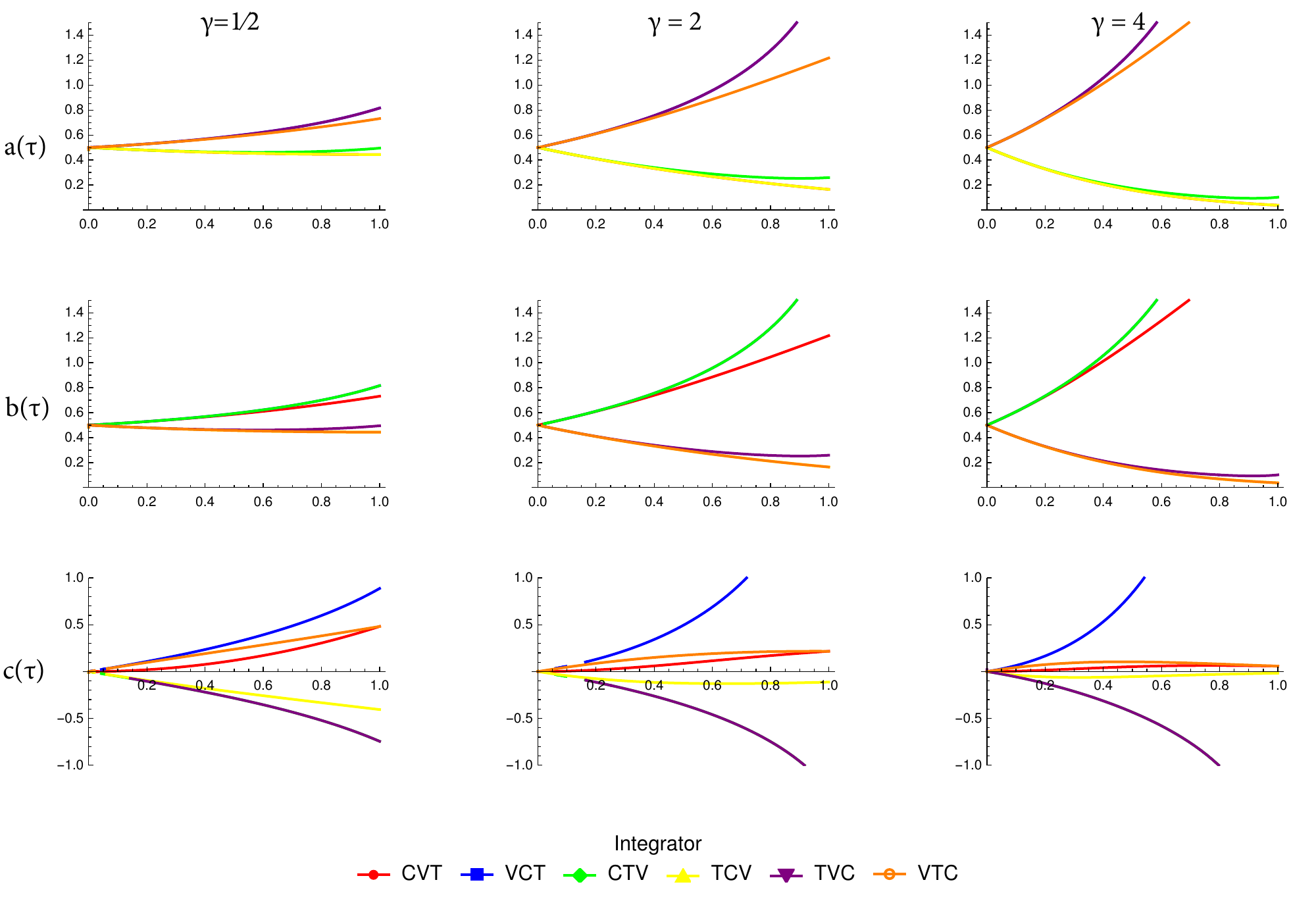}
  \caption{Plot of the parameters expressing the six possible modified Hamiltonians corresponding to the integrators in~\eqref{eq:1stintegrators}. 
  The rows are $a(\tau)$, $b(\tau)$ and $c(\tau)$, respectively, while along the columns we vary the parameter $\gamma \in \left\{\frac{1}{2}, 2, 4\right\}$.}
  \label{fig:parquaddump}
\end{figure}

As previously mentioned, the 1-dimensional damped harmonic oscillator belongs to the quadratic contact algebra $\Span_\mathbb{R}\{p^2,q^2,q\ p, s,{1}\}$ presented in the previous section. Indeed, its Hamiltonian is
\begin{equation}
	\label{eqn:hamdho}
    \cH = \frac{p^2}{2 m} + k \frac{q^2}{2} + \gamma s.
\end{equation}
In the rest of this section we will focus on the six possible 1st order integrators for the Hamiltonian~\eqref{eqn:hamdho}, that is,
\begin{align}
    S_{TVC}^\tau(x) &= \exp\{\tau X_{p^2/2}\} \exp\{ \tau X_{q^2/2}\} \exp\{\tau X_{\gamma s}\} \\
    S_{TCV}^\tau(x) &= \exp\{\tau X_{p^2/2}\} \exp\{ \tau X_{\gamma s}\} \exp\{\tau X_{q^2/2}\} \\
    S_{VCT}^\tau(x) &= \exp\{\tau X_{q^2/2}\} \exp\{ \tau X_{\gamma s}\} \exp\{\tau X_{p^2/2}\} \\ \label{eq:1stintegrators}
    S_{VTC}^\tau(x) &= \exp\{\tau X_{q^2/2}\} \exp\{\tau X_{p^2/2}\} \exp\{ \tau X_{\gamma s}\}\\
    S_{CTV}^\tau(x) &= \exp\{ \tau X_{\gamma s}\} \exp\{\tau X_{p^2/2}\} \exp\{\tau X_{q^2/2}\} \\
    S_{CVT}^\tau(x) &= \exp\{ \tau X_{\gamma s}\} \exp\{\tau X_{q^2/2}\} \exp\{\tau X_{p^2/2}\},
\end{align}
given by all the possible permutations of the three terms that form the Hamiltonian: the kinetic energy, the quadratic potential and the dissipation term. 
One can immediately check that their modified Hamiltonians must have the form
\begin{equation}
    \cH_\text{mod}(q,p,s;\tau) = a(\tau) q^2 + b(\tau) p^2 + {c}(\tau) q p + {d}(\tau) s,
\end{equation}
and equation~\eqref{eqns:deltazetacontquad} implies that $d(\tau) = \gamma$.
The remaining three parameters $a(\tau)$, $b(\tau)$ and $c(\tau)$ can be found from~\eqref{eq:tildeC}.
They
are plotted in Figure~\ref{fig:parquaddump} as functions of the timestep~$\tau$.

\begin{figure}[h]
  \centering
  \includegraphics[width=1.\linewidth]{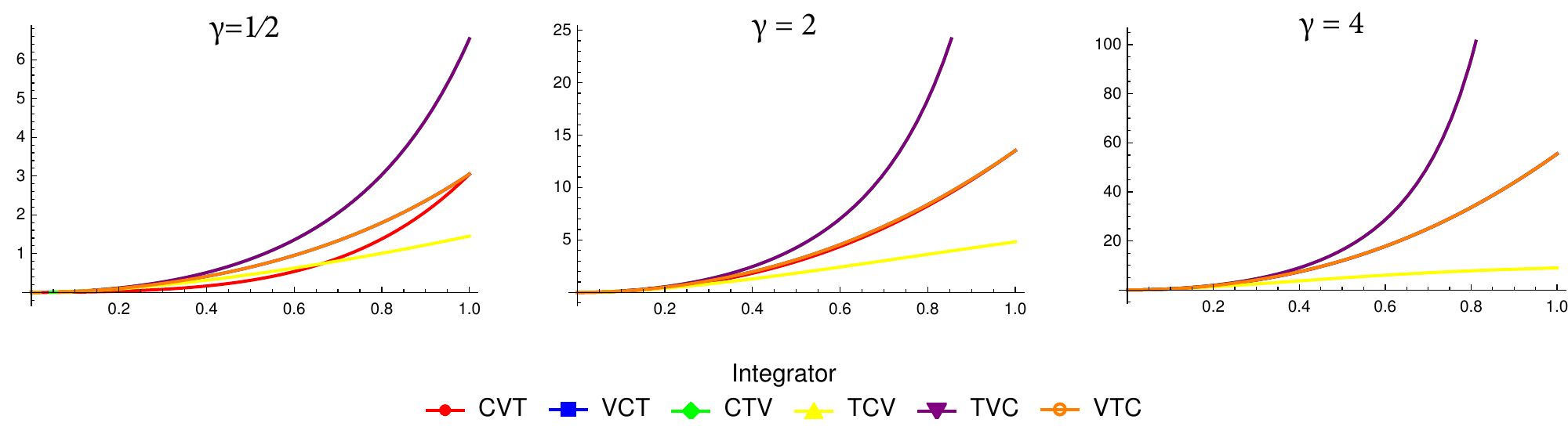}
  \caption{Plot of the trace distance between the original Hamiltonian and the six possible modified Hamiltonians
  corresponding to the integrators in~\eqref{eq:1stintegrators}.
  Along the columns we vary the parameter $\gamma\in\left\{\frac{1}{2}, 2, 4\right\}$.}
  \label{fig:parquaddumpham}
\end{figure}

In order to provide a quantitative comparison of the different modified Hamiltonians, 
we exploit the natural bilinear form induced on a Lie algebra 
by the \emph{Killing form}~\cite{Fulton1999}:
\begin{align}
    b      &:  \mathfrak{g} \times \mathfrak{g} \to \mathbb{R}\\
    b(X,Y) &:= {\rm Tr} (ad_X \circ ad _Y)\,,
\end{align}
where ${\rm Tr}$ is the trace operator and $ad_X$ is
the adjoint representation of the contact quadratic algebra
given in Appendix~\ref{app:matrixrepr}. 
Since $b(X,Y)$ in general fails to be positive-definite, we use $b(X,Y)^2$ as a (possibly degenerate) pseudo-distance on the space of contact Hamiltonian functions of the form~\eqref{eq:generalfuncitonquadratic}.
One can think of it as an analogue of the trace distance in quantum information theory and therefore we also refer to it as the \emph{trace distance} with a slight abuse of terminology.
In Figure~\ref{fig:parquaddumpham} we plot the trace distance between the original Hamiltonian and the modified Hamiltonian as a function of the time step $\tau$. Plotting the parameters and the distance from the original Hamiltonian already provides a visual way to select better-performing integrators: in almost all cases the integrator $S^\tau_{TCV}(x)$ has the least distance from the original Hamiltonian \eqref{eqn:hamdho}, and as such it is expected to perform better. At the same time, we can observe that for small values of $\gamma$, $S^\tau_{CVT}(x)$ could become a slightly better choice. Of course, at this stage, this is still a heuristic, but already hints at the potential usefulness of the method in the analysis of the performance of certain numerical integrators.

\section{Conclusions}\label{sec:Conclusions}

In this paper we presented a remarkably simple method, the flow method, to compute closed-form expressions for the Baker-Campbell-Hausdorff formula for Lie algebras that are finite-dimensional subalgebras of the Lie algebra of vector fields on an appropriate manifold.
In particular, we have focused on Lie algebras that can be expressed as algebras of contact Hamiltonian functions endowed with the Jacobi bracket.
Our original motivation for this investigation was to improve the error analysis of  contact splitting numerical integrators~\cite{Bravetti2020, Zadra_2021}, which we presented in the last section for the special example of the damped harmonic oscillator.
Having explicit modified Hamiltonians allows fine estimates on the numerical errors and a precise a priori comparison of the performance of different numerical algorithms.

Even though the flow method in its current form has only limited applicability (due to the fact that one needs to work with algebras of vector fields whose flows are integrable by quadratures), there are several possibilities for future research. 
We believe that the examples presented in the second part of the paper showcase its promises also in other contexts, for example it allowed us to recover in a remarkably simple way the result from \cite{Bravetti2017} and extend it to a larger class of algebras. {An adaptation of the techniques presented here to other Lie algebras is currently under investigation
and we expect to present the results in a following publication.}

The Baker-Campbell-Hausdorff formula is so pervasive in mathematics and physics~\cite{Bagarello_2022,Hall2015,Sakurai2011,Tuckerman2010} that it would not be surprising if more applications will emerge from different fields.
For instance, one could use the flow method in the context of the Kepler problem, after rewriting it as a harmonic oscillator~\cite{VANDERMEER2015181}, or for the numerical integration of time-dependent contact Hamiltonians. The latter should appear in a future work in the context of the numerical error analysis in the integration of contact Hamiltonians with time-dependent forcing and driving. {Moreover, squeezed states in quantum optics are quantum states constructed using the unitary representation of the symplectic group elements acting on the vacuum state of the quantum harmonic oscillator~\cite{gardiner2004quantum, braunstein2005quantum}. Our results could thus be used to enhance the analysis and the implementation of quantum gates, which can be considered as the product of two unitary symplectic operators~\cite{gardiner2004quantum, braunstein2005quantum}. They could also be extended to explore quantum gates in the context of polymer quantum mechanics once adapted to the polymer representation of the symplectic group~\cite{garcia2020symplectic, chacon2021relation, garcia2021propagation}. Moreover, the analysis and use of symplectic and contact integrators is a very active and prolific field e.g.~in mechanics~\cite{hairer2006geometric,Bravetti2020,anahory2021geometry,yoshida1993recent,Zadra_2021,esen2022discrete}, general relativity~\cite{wang2021construction1,wang2021construction2}
and plasma physics~\cite{qin2008variational, qin2015canonical},
and thus we consider that our results can be helpful in these contexts as well.}

\subsection*{Acknowledgments}
FZ would like to thank the IIMAS-UNAM 
for the kind hospitality during the preparation of this work. 
This visit was partially supported by NDNS+ project 2022.005. 
MS and FZ's research is supported by the NWO project 613.009.10.
The work of AB was partially supported by DGAPA-UNAM, program PAPIIT, Grant No. IA-102823.

\appendix

\section{Matrix representations}\label{app:matrixrepr}
An alternative (and equivalent) approach to compute the Baker-Campbell-Haussdorf formula by the flow method is to consider the matrix representation of the Lie algebra. In this Appendix we treat the examples in Section~\ref{sec:examples} from this point of view, giving particular attention to the adjoint representation.  

If we start from a Hamiltonian (either symplectic or contact) representation of a finite dimensional algebra~$\mathfrak{g}$, with generators the Hamiltonian functions $\{f_1, \cdots, f_n\}$, we can represent a general element of $\mathfrak{g}$ as the vector
\begin{equation}
  (\lambda_1, \cdots, \lambda_n) \sim \sum_i^n \lambda_i f_i\,.
\end{equation}
Exploiting the corresponding (Poisson or Jacobi) bracket $\{\cdot,\cdot\}$, we can associate to each element $a \in \mathfrak{g}$ the operator
\begin{equation}
  ad_a:=\{a, \cdot\} \,,
\end{equation}
which is called the \emph{adjoint operator 
corresponding to $a$}. 
Notice that $ad_a$ is a linear operator from $\mathfrak{g}$ to itself and therefore (for finite-dimensional algebras) 
it provides a way to associate to every element $a\in\mathfrak{g}$ a matrix (once a basis
of $\mathfrak{g}$ has been fixed).
The adjoint operator is always a morphism of Lie algebras, and thus provides a \emph{matrix representation} of $\mathfrak{g}$. 
This representation is particularly useful when $\mathfrak{g}$ is centerless, as shown by the following proposition~\cite{Fulton1999}.
\begin{Prop}\rm
  The adjoint representation is faithful, meaning that it is an injective morphism, if and only if the algebra is centerless.
\end{Prop}
If the algebra has a non-trivial center, then the adjoint representation is not faithful, since the center is mapped to the zero operator.
This is the case of the Heisenberg algebra in the symplectic representation, for which the center is the one-dimensional subspace generated by the Hamiltonian function $\cH=1$. On the contrary, the adjoint representation of the same algebra as a subalgebra of contact Hamiltonian functions is faithful.

\subsection{Matrix representation of the CHA}
\label{app:mrcha}
There is a representation of the contact Heisenberg  algebra~\cite{Bravetti2017} realised in the space of the real $4 \times 4$ matrices, provided by the following identifications:
\begin{align}
  q = \begin{pmatrix}
        0 & 0 & 0 & 0 \\
        0 & 0 & 1 & 0 \\
        0 & 0 & 0 & 1 \\
        0 & 0 & 0 & 0
      \end{pmatrix}, \qquad p = \begin{pmatrix}
                                  0 & 0 & -1 & 0 \\
                                  0 & 0 & 0  & 0 \\
                                  0 & 0 & 0  & 0 \\
                                  0 & 0 & 0  & 0
                                \end{pmatrix} \\
  s = \begin{pmatrix}
        0 & 0  & 0 & 0 \\
        0 & -1 & 0 & 0 \\
        0 & 0  & 0 & 0 \\
        0 & 0  & 0 & 1
      \end{pmatrix} \qquad {1} = \begin{pmatrix}
                                          0 & 0 & 0 & 1 \\
                                          0 & 0 & 0 & 0 \\
                                          0 & 0 & 0 & 0 \\
                                          0 & 0 & 0 & 0
                                        \end{pmatrix}.
\end{align}
Indeed, the commutators of these matrices fulfill the contact Heisenberg relations
\begin{equation}
\label{eqn:commrelCHA}
  [q,p]={1}, \quad [{1},s]= {1},
  \quad \text{and} \quad [p,s] = {1}\,,
\end{equation}
where $[a,b]:=ab-ba$ 
and all the remaining commutators vanish.

Another representation is the \emph{adjoint representation}, which can be found as described in the previous section by exploiting the Jacobi bracket. 
It leads to the new form of the commutation relations \eqref{eqn:commrelCHA}
\begin{equation}
  \label{eqn:firstcomm}
  \left\{(z,a,b,c),\ (\bar{z},\bar{a},\bar{b},\bar{c})\right\}_\eta = (z_1,a_1,b_1, c_1 ),
\end{equation}
where \cite{Bravetti2017}
\begin{align}%
  z_1 & = +\bar{c} z - c \bar{z} + \bar{a} b - \bar{b} a, \\
  a_1 & = -\bar{a} c + a \bar{c} ,                        \\
  b_1 & = 0,                                              \\
  c_1 & = 0.
\end{align}
We can use this to extract the matrix representation of the adjoint since
\begin{equation}
  M_f \cdot (\bar{z},\bar{a},\bar{b},\bar{c}) = (z_1,a_1,0,0 ),
\end{equation}
that is,
\begin{equation}
  \label{eq:secondCCHAamtrices}
  M_f = \begin{pmatrix} -c & b  & -a & z \\
                0  & -c & 0  & a \\
                0  & 0  & 0  & 0 \\
                0  & 0  & 0  & 0\end{pmatrix}.
\end{equation}
For the sake of simplicity we present only the computations using the first representation, 
but the results are equivalent if one uses the representation~\eqref{eq:secondCCHAamtrices}.

For a contact Hamiltonian in the CHA algebra,
\begin{equation}
  f = a q + b p + c s + z {1},
\end{equation}
the matrix exponential $e^f$ is explicitly computed as
\begin{equation}
  e^f = \begin{pmatrix}
    1 & 0      & -b                                    & \frac{-a b e^c+a b c+a b+c e^c z-c z}{c^2}    \\
    0 & e^{-c} & \frac{a e^{-c} \left(e^c-1\right)}{c} & \frac{a^2 e^{-c} \left(e^c-1\right)^2}{2 c^2} \\
    0 & 0      & 1                                     & \frac{a \left(e^c-1\right)}{c}                \\
    0 & 0      & 0                                     & e^c
  \end{pmatrix}.
\end{equation}
We do not need to explicitly compute the matrix logarithm of the product of two such exponentials and instead we can work directly with the group elements: since the aforementioned matrix logarithm is again an element of the algebra, the coefficients $\alpha, \beta, \gamma$ and $\zeta$ in
\begin{equation}
  \mathcal{f} = \alpha q + \beta p + \gamma s + \zeta {1}
\end{equation}
can be identified by equating term by term, immediately leading to \eqref{eqn:solutioncontactheisemberg}.

\subsection{Matrix representation of QCA}
\label{app:mrqca}
One can in principle repeat the steps presented above to find a representation of the Quadratic Contact Algebra (see Section \ref{sec:QCA}).
The commutator between two general elements of the algebra \eqref{eqn:firstcomm} can be written in the form
\begin{equation}
  \left\{(z,a,b,c,d),\ (\bar{z},\bar{a},\bar{b},\bar{c},\bar{d})\right\}_\eta = (z_1,a_1,b_1, c_1 , 0 ),
\end{equation}
where
\begin{align}%
  z_1 & = +\bar{d} z - d \bar{z},                             \\
  a_1 & = -\bar{a} d + a \bar{d} - 2 \bar{a} c + 2 a \bar{c}, \\
  b_1 & = \bar{b} d - b \bar{d} + 2 \bar{b} c - 2 b \bar{c},  \\
  c_1 & = -4 \bar{a} b + 4 a \bar{b}                          \\
  d_1 & = 0.
\end{align}
In this case, the matrix representation of the adjoint is
\begin{equation}
  \begin{pmatrix}
    -d & 0      & 0     & 0    & z  \\
    0  & -2 c-d & 0     & 2 a  & a  \\
    0  & 0      & 2 c+d & -2 b & -b \\
    0  & -4 b   & 4 a   & 0    & 0  \\
    0  & 0      & 0     & 0    & 0  \\
  \end{pmatrix},
\end{equation}
and the generators of the algebra are
\begin{align}
  q^2 = \begin{pmatrix}
          0 & 0 & 0 & 0 & 0 \\
          0 & 0 & 0 & 2 & 1 \\
          0 & 0 & 0 & 0 & 0 \\
          0 & 0 & 4 & 0 & 0 \\
          0 & 0 & 0 & 0 & 0 \\
        \end{pmatrix} & \qquad p^2 = (-) \begin{pmatrix}
                                           0 & 0 & 0 & 0 & 0 \\
                                           0 & 0 & 0 & 0 & 0 \\
                                           0 & 0 & 0 & 2 & 1 \\
                                           0 & 4 & 0 & 0 & 0 \\
                                           0 & 0 & 0 & 0 & 0 \\
                                         \end{pmatrix} \\
  q p=\begin{pmatrix}
        0 & 0  & 0 & 0 & 0 \\
        0 & -2 & 0 & 0 & 0 \\
        0 & 0  & 2 & 0 & 0 \\
        0 & 0  & 0 & 0 & 0 \\
        0 & 0  & 0 & 0 & 0 \\
      \end{pmatrix} \qquad &
  s = \begin{pmatrix}
        -1 & 0  & 0 & 0 & 0 \\
        0  & -1 & 0 & 0 & 0 \\
        0  & 0  & 1 & 0 & 0 \\
        0  & 0  & 0 & 0 & 0 \\
        0  & 0  & 0 & 0 & 0 \\
      \end{pmatrix} \qquad 
 {1} = \begin{pmatrix}
          0 & 0 & 0 & 0 & 1 \\
          0 & 0 & 0 & 0 & 0 \\
          0 & 0 & 0 & 0 & 0 \\
          0 & 0 & 0 & 0 & 0 \\
          0 & 0 & 0 & 0 & 0 \\
        \end{pmatrix}.
\end{align}
In this case, however, it is much easier to apply the flow method: the computation of the matrix exponential of a $5\times 5$ matrix is much more cumbersome.
One can already foresee how, with larger algebras, this difference will start getting larger and the flow method could become increasingly advantageous.

\nocite{*}
\printbibliography

@Control{biblatex-control,
  options = {3.7:0:0:1:0:1:1:0:0:0:0:1:3:1:3:1:0:0:3:1:79:+:+:nty},
}

@Article{Bravetti2017,
  author    = {Alessandro Bravetti and Angel Garcia-Chung and Diego Tapias},
  journal   = {Journal of Physics A: Mathematical and Theoretical},
  title     = {Exact Baker--Campbell--Hausdorff formula for the contact Heisenberg algebra},
  year      = {2017},
  month     = {feb},
  number    = {10},
  pages     = {105203},
  volume    = {50},
  doi       = {10.1088/1751-8121/aa59dd},
  publisher = {{IOP} Publishing},
}

@Article{Bravetti2020,
  author    = {Alessandro Bravetti and Marcello Seri and Mats Vermeeren and Federico Zadra},
  journal   = {Celestial Mechanics and Dynamical Astronomy},
  title     = {Numerical integration in Celestial Mechanics: a case for contact geometry},
  year      = {2020},
  month     = {jan},
  number    = {1},
  volume    = {132},
  doi       = {10.1007/s10569-019-9946-9},
  publisher = {Springer Science and Business Media {LLC}},
}

@Article{Kirillov1976,
  author    = {A. A. Kirillov},
  journal   = {Russian Mathematical Surveys},
  title     = {Local Lie Algebras},
  year      = {1976},
  month     = {aug},
  number    = {4},
  pages     = {55--75},
  volume    = {31},
  doi       = {10.1070/rm1976v031n04abeh001556},
  publisher = {{IOP} Publishing},
}

@Book{Hall2015,
  author    = {Hall, Brian},
  publisher = {Springer-Verlag GmbH},
  title     = {Lie Groups, Lie Algebras, and Representations},
  year      = {2015},
  isbn      = {3319134663},
  month     = may,
  ean       = {9783319134666},
}

@Article{Bravetti2017a,
  author    = {Alessandro Bravetti and Hans Cruz and Diego Tapias},
  journal   = {Annals of Physics},
  title     = {Contact Hamiltonian mechanics},
  year      = {2017},
  month     = {jan},
  pages     = {17--39},
  volume    = {376},
  doi       = {10.1016/j.aop.2016.11.003},
  publisher = {Elsevier {BV}},
}

@Article{Bravetti2017b,
  author    = {Alessandro Bravetti},
  journal   = {Entropy},
  title     = {Contact Hamiltonian Dynamics: The Concept and Its Use},
  year      = {2017},
  month     = {oct},
  number    = {10},
  pages     = {535},
  volume    = {19},
  doi       = {10.3390/e19100535},
  publisher = {{MDPI} {AG}},
}

@Article{Bravetti2019,
  author    = {Alessandro Bravetti},
  journal   = {International Journal of Geometric Methods in Modern Physics},
  title     = {Contact geometry and thermodynamics},
  year      = {2019},
  month     = {jan},
  number    = {supp01},
  pages     = {1940003},
  volume    = {16},
  doi       = {10.1142/s0219887819400036},
  publisher = {World Scientific Pub Co Pte Lt},
}

@Article{Yoshida1990,
  author    = {Haruo Yoshida},
  journal   = {Physics Letters A},
  title     = {Construction of higher order symplectic integrators},
  year      = {1990},
  month     = {nov},
  number    = {5-7},
  pages     = {262--268},
  volume    = {150},
  doi       = {10.1016/0375-9601(90)90092-3},
  publisher = {Elsevier {BV}},
}

@Book{Varadarajan1974,
  author    = {Varadarajan, V. S.},
  publisher = {Prentice-Hall},
  title     = {Lie groups, Lie algebras, and their representations},
  year      = {1974},
  address   = {Englewood Cliffs, N.J},
  isbn      = {9780135357323},
}

@Book{P.Libermann1987,
  author    = {P. Libermann, Charles-Michel Marle},
  publisher = {Springer Netherlands},
  title     = {Symplectic Geometry and Analytical Mechanics},
  year      = {1987},
  isbn      = {9027724385},
  month     = mar,
  ean       = {9789027724380},
  pagetotal = {546},
}

@article{Zadra_2021,
	doi = {10.3390/math9161960},
	year = 2021,
	month = {aug},
	publisher = {{MDPI} {AG}},
	volume = {9},
	number = {16},
	pages = {1960},
	author = {Federico Zadra and Alessandro Bravetti and Marcello Seri},
	title = {Geometric Numerical Integration of Li{\'{e}}nard Systems via a Contact Hamiltonian Approach},
	journal = {Mathematics}
}

@book{arnol2013mathematical,
  title={Mathematical methods of classical mechanics},
  author={Arnol'd, Vladimir Igorevich},
  volume={60},
  year={2013},
  publisher={Springer Science \& Business Media}
}

@Article{Donnelly2005,
  author    = {Denis Donnelly and Edwin Rogers},
  journal   = {American Journal of Physics},
  title     = {Symplectic integrators: An introduction},
  year      = {2005},
  month     = {oct},
  number    = {10},
  pages     = {938--945},
  volume    = {73},
  doi       = {10.1119/1.2034523},
  publisher = {American Association of Physics Teachers ({AAPT})},
}

@Book{Fasano2002,
  author    = {Fasano, A.},
  publisher = {Oxford University Press},
  title     = {Analytical mechanics},
  year      = {2002},
  isbn      = {9780198508021},
  subtitle  = {{A}n introduction},
}

@Article{Boyer2011,
  author    = {Charles P. Boyer},
  journal   = {Symmetry, Integrability and Geometry: Methods and Applications},
  title     = {Completely Integrable Contact {H}amiltonian Systems and Toric Contact Structures on $S^2\times S^3$},
  year      = {2011},
  month     = {jun},
  doi       = {10.3842/sigma.2011.058},
  publisher = {{SIGMA} (Symmetry, Integrability and Geometry: Methods and Application)},
}

@Book{Fulton1999,
  author    = {Fulton, William and Harris, Joe},
  publisher = {Springer},
  title     = {Representation Theory},
  year      = {1999},
  isbn      = {9780387974958},
  pages     = {551},
  subtitle  = {A First Course (Graduate Texts in Mathematics / Readings in Mathematics)},
}

@Article{Skeel2001,
  author    = {Skeel, Robert and Hardy, David},
  journal   = {SIAM Journal on Scientific Computing},
  title     = {Practical Construction of Modified Hamiltonians},
  year      = {2001},
  month     = {1},
  number    = {4},
  pages     = {1172-1188},
  volume    = {23},
  date      = {2001-01},
  doi       = {10.1137/s106482750138318x},
  publisher = {Society for Industrial & Applied Mathematics (SIAM)},
}

@Article{Biagi2013a,
  author    = {Stefano Biagi and Andrea Bonfiglioli},
  journal   = {Linear and Multilinear Algebra},
  title     = {On the convergence of the Campbell--Baker--Hausdorff--Dynkin series in infinite-dimensional Banach-Lie algebras},
  year      = {2013},
  month     = {10},
  number    = {12},
  pages     = {1591-1615},
  volume    = {62},
  date      = {2013-10-11},
  day       = {11},
  doi       = {10.1080/03081087.2013.839674},
  publisher = {Informa UK Limited},
}

@Article{Weigert1997,
  author    = {Weigert, Stefan},
  journal   = {Journal of Physics A: Mathematical and General},
  title     = {Baker-Campbell-Hausdorff relation for special unitary groups},
  year      = {1997},
  month     = {12},
  number    = {24},
  pages     = {8739-8749},
  volume    = {30},
  date      = {1997-12-21},
  day       = {21},
  doi       = {10.1088/0305-4470/30/24/032},
  publisher = {IOP Publishing},
}

@Article{Matone_JHEP_2015,
  author    = {Marco Matone},
  journal   = {Journal of High Energy Physics},
  title     = {An algorithm for the Baker-Campbell-Hausdorff formula},
  year      = {2015},
  month     = {may},
  number    = {5},
  volume    = {2015},
  doi       = {10.1007/jhep05(2015)113},
  publisher = {Springer Science and Business Media {LLC}},
}

@Article{Matone_JGP_2015,
  author    = {Marco Matone},
  journal   = {Journal of Geometry and Physics},
  title     = {Classification of commutator algebras leading to the new type of closed Baker-Campbell-Hausdorff formulas},
  year      = {2015},
  month     = {nov},
  pages     = {34--43},
  volume    = {97},
  doi       = {10.1016/j.geomphys.2015.06.016},
  publisher = {Elsevier {BV}},
}

@Article{Van_Brunt_2018,
  author    = {Alexander Van-Brunt and Matt Visser},
  journal   = {Mathematics},
  title     = {Explicit Baker-Campbell-Hausdorff Expansions},
  year      = {2018},
  month     = {aug},
  number    = {8},
  pages     = {135},
  volume    = {6},
  doi       = {10.3390/math6080135},
  publisher = {{MDPI} {AG}},
}

@article{Liu_2021,
	doi = {10.3934/dcdsb.2021297},
	year = 2021,
	publisher = {American Institute of Mathematical Sciences ({AIMS})},
	author = {Qihuai Liu and Pedro J. Torres},
	title = {Orbital dynamics on invariant sets of contact Hamiltonian systems},
	journal = {Discrete {\&} Continuous Dynamical Systems - B}
}

@Article{Achilles2012,
  author    = {Rüdiger Achilles and Andrea Bonfiglioli},
  journal   = {Archive for History of Exact Sciences},
  title     = {The early proofs of the theorem of Campbell, Baker, Hausdorff, and Dynkin},
  year      = {2012},
  month     = {apr},
  number    = {3},
  pages     = {295--358},
  volume    = {66},
  doi       = {10.1007/s00407-012-0095-8},
  publisher = {Springer Science and Business Media {LLC}},
}

@InCollection{Bagarello_2022,
  author    = {Fabio Bagarello},
  booktitle = {Pseudo-Bosons and Their Coherent States},
  publisher = {Springer International Publishing},
  title     = {Our Way to the {BCH} Formula},
  year      = {2022},
  pages     = {71--88},
  doi       = {10.1007/978-3-030-94999-0_4},
}

@Article{de_Le_n_2019,
  author    = {Manuel de Le{\'{o}}n and Manuel Lainz Valc{\'{a}}zar},
  journal   = {Journal of Mathematical Physics},
  title     = {Contact Hamiltonian systems},
  year      = {2019},
  month     = {oct},
  number    = {10},
  pages     = {102902},
  volume    = {60},
  doi       = {10.1063/1.5096475},
  publisher = {{AIP} Publishing},
}

@Book{Tuckerman2010,
  author    = {Tuckerman, Mark E.},
  publisher = {Oxford University Press},
  title     = {Statistical mechanics},
  year      = {2010},
  isbn      = {9780198525264},
  pages     = {696},
  subtitle  = {theory and molecular simulation},
}

@Book{Sakurai2011,
  author    = {Sakurai, J. J.},
  publisher = {Addison Wesley},
  title     = {Modern quantum mechanics - 2. Edition},
  year      = {2011},
  isbn      = {9780805382914},
}

@article{AIHPA_1970__13_2_103_0,
     author = {Tilgner, Hans},
     title = {A class of solvable {Lie} groups and their relation to the canonical formalism},
     journal = {Annales de l'I.H.P. Physique th\'eorique},
     pages = {103--127},
     publisher = {Gauthier-Villars},
     volume = {13},
     number = {2},
     year = {1970},
     mrnumber = {277192},
}

@article{VANDERMEER2015181,
title = {The Kepler system as a reduced 4D harmonic oscillator},
journal = {Journal of Geometry and Physics},
volume = {92},
pages = {181-193},
year = {2015},
issn = {0393-0440},
doi = {10.1016/j.geomphys.2015.02.016},
author = {J.C. {van der Meer}},
}

@article{Albert_1989,
	doi = {10.1016/0393-0440(89)90029-6},
	url = {https://doi.org/10.1016/0393-0440(89)90029-6},
	year = 1989,
	publisher = {Elsevier {BV}},
	volume = {6},
	number = {4},
	pages = {627--649},
	author = {Claude Albert},
	title = {Le th{\'{e}}or{\`{e}}me de r{\'{e}}duction de Marsden-Weinstein en g{\'{e}}om{\'{e}}trie cosymplectique et de contact},
	journal = {Journal of Geometry and Physics}
}

@book{gardiner2004quantum,
  title={Quantum noise: a handbook of Markovian and non-Markovian quantum stochastic methods with applications to quantum optics},
  author={Gardiner, Crispin and Zoller, Peter and Zoller, Peter},
  year={2004},
  publisher={Springer Science \& Business Media}
}

@article{braunstein2005quantum,
  title={Quantum information with continuous variables},
  author={Braunstein, Samuel L and Van Loock, Peter},
  journal={Reviews of modern physics},
  volume={77},
  number={2},
  pages={513},
  year={2005},
  publisher={APS}
}

@article{garcia2020symplectic,
  title={Symplectic group in polymer quantum mechanics},
  author={Garcia-Chung, Angel},
  journal={Physical Review D},
  volume={101},
  number={10},
  pages={106004},
  year={2020},
  publisher={APS}
}

@article{chacon2021relation,
  title={The relation between the symplectic group $Sp(4, R)$ and its Lie algebra: Applications to polymer quantum mechanics},
  author={Chac{\'o}n-Acosta, Guillermo and Garcia-Chung, Angel},
  journal={Physical Review D},
  volume={104},
  number={12},
  pages={126006},
  year={2021},
  publisher={APS}
}

@article{garcia2021propagation,
  title={Propagation of quantum gravity-modified gravitational waves on a classical FLRW spacetime},
  author={Garcia-Chung, Angel and Mertens, James B and Rastgoo, Saeed and Tavakoli, Yaser and Moniz, Paulo Vargas},
  journal={Physical Review D},
  volume={103},
  number={8},
  pages={084053},
  year={2021},
  publisher={APS}
}

@article{yoshida1993recent,
  title={Recent progress in the theory and application of symplectic integrators},
  author={Yoshida, Haruo},
  journal={Qualitative and Quantitative Behaviour of Planetary Systems},
  pages={27--43},
  year={1993},
  publisher={Springer}
}

@article{qin2008variational,
  title={Variational symplectic integrator for long-time simulations of the guiding-center motion of charged particles in general magnetic fields},
  author={Qin, Hong and Guan, Xiaoyin},
  journal={Physical review letters},
  volume={100},
  number={3},
  pages={035006},
  year={2008},
  publisher={APS}
}

@article{qin2015canonical,
  title={Canonical symplectic particle-in-cell method for long-term large-scale simulations of the Vlasov--Maxwell equations},
  author={Qin, Hong and Liu, Jian and Xiao, Jianyuan and Zhang, Ruili and He, Yang and Wang, Yulei and Sun, Yajuan and Burby, Joshua W and Ellison, Leland and Zhou, Yao},
  journal={Nuclear Fusion},
  volume={56},
  number={1},
  pages={014001},
  year={2015},
  publisher={IOP Publishing}
}

@article{wang2021construction1,
  title={Construction of Explicit Symplectic Integrators in General Relativity. {I}. {S}chwarzschild Black Holes},
  author={Wang, Ying and Sun, Wei and Liu, Fuyao and Wu, Xin},
  journal={The Astrophysical Journal},
  volume={907},
  number={2},
  pages={66},
  year={2021},
  publisher={IOP Publishing}
}

@article{gaset2020new,
  title={New contributions to the {H}amiltonian and {L}agrangian contact formalisms for dissipative mechanical systems and their symmetries},
  author={Gaset, Jordi and Gr{\`a}cia, Xavier and Mu{\~n}oz-Lecanda, Miguel C and Rivas, Xavier and Rom{\'a}n-Roy, Narciso},
  journal={International Journal of Geometric Methods in Modern Physics},
  volume={17},
  number={06},
  pages={2050090},
  year={2020},
  publisher={World Scientific}
}

@article{grabowska2022novel,
  title={A novel approach to contact {H}amiltonians and contact {H}amilton-{J}acobi theory},
  author={Grabowska, Katarzyna and Grabowski, Janusz},
  journal={arXiv preprint arXiv:2207.04484},
  year={2022}
}

@article{anahory2021geometry,
  title={On the geometry of discrete contact mechanics},
  author={Anahory Simoes, Alexandre and Mart{\'\i}n de Diego, David and Lainz Valc{\'a}zar, Manuel and de Le{\'o}n, Manuel},
  journal={Journal of Nonlinear Science},
  volume={31},
  number={3},
  pages={1--30},
  year={2021},
  publisher={Springer}
}

@book{hairer2006geometric,
  publisher = {Springer Berlin, Heidelberg},
  isbn      = {978-3-540-30663-4},
  title={Geometric numerical integration. Structure-Preserving Algorithms for Ordinary Differential Equations},
  author={Hairer, Ernst and Wanner, Gerhard and Lubich, Christian},
  year={2006}
}

@article{esen2022discrete,
  title={A Discrete {H}amilton--{J}acobi Theory for Contact {H}amiltonian Dynamics},
  author={Esen, O{\u{g}}ul and Sard{\'o}n, Cristina and Zajac, Marcin},
  journal={arXiv preprint arXiv:2209.05922},
  year={2022}
}

@article{wang2021construction2,
  title={Construction of Explicit Symplectic Integrators in General Relativity. {II}. {R}eissner--{N}ordstr{\"o}m Black Holes},
  author={Wang, Ying and Sun, Wei and Liu, Fuyao and Wu, Xin},
  journal={The Astrophysical Journal},
  volume={909},
  number={1},
  pages={22},
  year={2021},
  publisher={IOP Publishing}
}

@article{bravetti2022scaling,
  title={Scaling Symmetries, Contact Reduction and {P}oincar\'e's dream},
  author={Bravetti, Alessandro and Jackman, Connor and Sloan, David},
  journal={arXiv preprint arXiv:2206.09911},
  year={2022}
}

@book{leimkuhler2004simulating,
  title={Simulating {H}amiltonian dynamics},
  author={Leimkuhler, Benedict and Reich, Sebastian},
  number={14},
  year={2004},
  publisher={Cambridge university press}
}

\end{document}